%% file: LineSurveysFullPostProofsText.tex
\documentclass[12pt,preprint]{aastex}

\usepackage{natbib}
\begin{document}
\def\phb#1{#1} 
\def\reac#1#2{\begin{equation}\label{r:r#1}\ce{#2}\end{equation}}
\def\reacnn#1{\[\ce{#1}\]}
\def\refreac#1{(\ref{r:r#1})}
\def\abS{\ensuremath{[\ce{S}]_{\rm tot}}}
\def\ab#1{\ensuremath{[\ce{#1}]}}
\def\fch2{\ensuremath{f_{\ce{CH2}}}}
\def\fhcn{\ensuremath{f_{\rm HCN}}}
\def\fhnc{\ensuremath{f_{\rm HNC}}}
\def\fcn{\ensuremath{f_{\rm CN}}}
\def\fifn{\ensuremath{\ce{^{15}N}}}
\def\fif{$^{15}$}
\def\fifnp{\ensuremath{\ce{^{15}N+}}}
\def\foun{\ensuremath{\ce{$^{14}$N}}}
\def\fifno{\ensuremath{\ce{$^{15}$NO}}}
\def\cfifn{\ensuremath{\ce{C$^{15}$N}}}
\def\nfifn{\ensuremath{\ce{N$^{15}$N}}}
\def\cfoun{\ensuremath{\ce{C$^{14}$N}}}
\def\hcfifn{\ensuremath{\ce{HC$^{15}$N}}}
\def\hfifnc{\ensuremath{\ce{H$^{15}$NC}}}
\def\conv{\ensuremath{{\cal N}\ul}}
\def\tauul{\ensuremath{\tau_{ul}}}
\def\ul{\ensuremath{{_{ul}}}}
\def\tul{\ensuremath{T_{ul}}}
\def\aul{\ensuremath{A_{ul}}}
\def\gup{\ensuremath{g_u}}
\def\Ntot{\ensuremath{N_{\rm tot}}}
\def\qpart{\ensuremath{Q_{\rm exc}}}
\def\qtot{\ensuremath{Q_{\rm tot}}}
\def\qrot{\ensuremath{Q_{\rm rot}}}
\def\rr{\ensuremath{\thcn/\cfifn}}
\def\rrr{\ensuremath{\cfoun/\cfifn}}
\def\nratio {\ensuremath{{\foun/\fifn}}}
\def\nhhd   {\ensuremath{\rm NH_2D}}
\def\nnhp   {\ensuremath{\rm N_2H^+}}
\def\NH    {\ensuremath{\rm NH}}
\def\NHH    {\ensuremath{\rm NH_2}}
\def\nhhh   {\ensuremath{\rm NH_3}}
\def\fifnhh {\ensuremath{\rm ^{15}NH_2}}
\def\fifnhhh{\ensuremath{\rm ^{15}NH_3}}
\def\fifnnhp{\ensuremath{\rm ^{15}NNH^+}}
\def\nfifnhp{\ensuremath{\rm N^{15}NH^+}}
\def\perth{\ensuremath{\hbox{$\,^0\!/_{00}$}}}
\def\perth{\textperthousand}
\def\jnu{\ensuremath{J_\nu}}
\def\tpeak{\ensuremath{T_{\rm mb, 0}}}
\def\tpeak{\ensuremath{T_0}}
\def\aff{\ensuremath{A_{FF'}}}
\def\fbeam{\ensuremath{f_{\rm dilution}}}

\title{
Unbiased mm-wave Line Surveys of TW~Hya and V4046~Sgr: \\ The Enhanced C$_2$H and CN Abundances of Evolved Protoplanetary Disks} 

\author{Joel H.\ Kastner\altaffilmark{1}, Pierre Hily-Blant\altaffilmark{2}, David R. Rodriguez\altaffilmark{3}, Kristina Punzi\altaffilmark{1}, Thierry Forveille\altaffilmark{2}}


\altaffiltext{1}{Center for Imaging Science, School of Physics \& Astronomy, and 
Laboratory for Multiwavelength Astrophysics, Rochester Institute of
Technology, 54 Lomb Memorial Drive, Rochester NY 14623 USA
(jhk@cis.rit.edu)}
\altaffiltext{2}{UJF-Grenoble 1/CNRS-INSU, Institut de Plan\'{e}tologie et d'Astrophysique de Grenoble (IPAG) UMR 5274, 38041, Grenoble, France}
\altaffiltext{3}{Departamento de Astronom\'ia, Universidad de Chile, Casilla 36-D, Santiago, Chile}

\begin{abstract}
We have conducted the first comprehensive mm-wave molecular emission line surveys of the evolved circumstellar disks orbiting the nearby, roughly solar-mass, pre-main sequence (T Tauri) stars TW Hya ($D = 54$ pc) and V4046~Sgr~AB ($D = 73$ pc). Both disks are known to retain significant residual gaseous components, despite the advanced ages of their host stars ($\sim$8 Myr and $\sim$21 Myr, respectively). Our unbiased broad-band radio spectral surveys of the TW Hya and V4046 Sgr disks were performed with the Atacama Pathfinder Experiment (APEX) 12 meter telescope and are intended to yield a complete census of bright molecular emission lines in the range 275--357 GHz (1.1--0.85 mm).  We find that lines of $^{12}$CO, $^{13}$CO, HCN, CN, and C$_2$H, all of which lie in the higher-frequency ($>$330 GHz) range, constitute the strongest molecular emission from both disks in the spectral region surveyed. 
The molecule C$_2$H is detected here for the first time in both disks, as is CS in the TW Hya disk. 
The survey results also include the first measurements of the full suite of hyperfine transitions of CN $N=3\rightarrow2$ and C$_2$H $N=4\rightarrow3$ in both disks. Modeling of these CN and C$_2$H hyperfine complexes in the spectrum of TW~Hya indicates that the emission from both species is optically thick and may originate from very cold ($\stackrel{<}{\sim}$10 K) disk regions. The latter result, if confirmed, would suggest efficient production of CN and C$_2$H in the outer disk and/or near the disk midplane. It furthermore appears that the fractional abundances of CN and C$_2$H are significantly enhanced in these evolved protoplanetary disks relative to the fractional abundances of the same molecules in the environments of deeply embedded protostars. These results, combined with previous determinations of enhanced abundances of other species (such as HCO$^+$) in T Tauri star disks, underscore the importance of properly accounting for high-energy (FUV and X-ray) radiation from the central T Tauri star when modeling protoplanetary disk gas chemistry and physical conditions. 
\end{abstract}

\section{Introduction}

Millimeter and submm interferometric studies of molecular emission from the nearest and most evolved known examples of dusty disks orbiting pre-main sequence (pre-MS) stars provide a unique means to ascertain the physical conditions and chemistry within outer ($R>10$ AU) disk regions \citep[see reviews in][]{2011ARA&A..49...67W,2014arXiv1402.3503D}. In evolved disks with ages of a few Myr, these regions are likely undergoing the buildup and orbital migration of giant planets and icy planetesimals \citep[e.g.,][]{2002ARA&A..40..103H,2002ARA&A..40...63L}. Submillimeter molecular line observations thereby serve to clarify the ``voyage'' of organic material from interstellar clouds to gaseous and icy outer solar system bodies \citep{2000ARA&A..38..427E}. 

Theorists have increasingly focused on pre-MS star UV and X-ray radiation as potential drivers of the processing of planet-forming gas and dust in disks, via the profound effects of such radiation on disk gas heating and chemistry \citep[e.g.,][]{1997ApJ...480..344G,2004ApJ...615..972G,2009ApJ...701..142G,2012ApJ...756..157G,2005A&A...440..949S,2008ApJ...683..287G,2008ApJ...676..518M,2008ApJ...688..398E,2009ApJ...699.1639E,2009ApJ...699L..35D,2009ApJ...705.1237G,2011ApJ...735...90G,2010MNRAS.401.1415O,2011MNRAS.412...13O,2012ApJ...747..114W,2012A&A...547A..69A,2013ApJ...772....5C}.
In parallel, we and others are undertaking submillimeter observations of potential molecular tracers of disk irradiation, so as to inform and test such models \citep[e.g.,][]{2008A&A...492..469K,2008ApJ...681.1396Q,2010ApJ...714.1511H,2011A&A...536A..80S,2011ApJ...734...98O,2012A&A...537A..60C}.

At distances of only 54 and 73 pc, respectively \citep[][and refs.\ therein]{2008hsf2.book..757T}, the relatively old (age $\sim$10 Myr) yet still actively accreting pre-MS star/disk systems TW Hya and V4046 Sgr offer unparalleled opportunities to study late-stage disk dissipation and planet-building processes at close range. Given their roughly solar-mass central stars, these circumstellar environments are likely similar to that of the early solar system --- with the important and intriguing caveat that TW Hya is a single star of mass $\sim$0.8 $M_\odot$ \citep{2012ApJ...744..162A}, whereas V4046 Sgr is a close ($P\sim 2.4$ day) binary system consisting of nearly equal-mass, $\sim$0.9 $M_\odot$ components \citep{2012ApJ...759..119R}. Despite their advanced ages \citep[estimated as $\sim$8 Myr and $\sim$21 Myr, respectively;][]{2008hsf2.book..757T,2014MNRAS.438L..11B,2014A&A...563A.121D}, the circumstellar disks orbiting TW Hya and V4046 Sgr both  retain substantial reservoirs of molecular gas --- according to some estimates, as much as $\sim$0.05--0.1 $M_\odot$ of H$_2$ --- as evidenced by relatively strong emission lines of CO and other trace molecular species detected and mapped thus far \citep[e.g.,][]{1997Sci...277...67K,2008A&A...492..469K,2004A&A...425..955T,2004ApJ...616L..11Q,2006ApJ...636L.157Q,2008ApJ...681.1396Q,2010ApJ...720.1684R,2011ApJ...734...98O,2012ApJ...744..162A,2012ApJ...757..129R,2012ApJ...759..119R,2013ApJ...775..136R,2013Natur.493..644B}.
These results have established the importance of the disks orbiting TW Hya and V4046 Sgr to studies of disk molecular chemistry and disk dissipation processes \citep[for a summary of the properties of these two star/disk systems, see Table 3 in ][]{2014A&A...561A..42S}.
In particular, submm molecular line observations of these two disks suggest that the abundances of certain species --- such as HCN, CN, and HCO$^+$, all of which are well-detected in these and other, similarly evolved disks orbiting active, late-type stars --- are enhanced by strong UV and X-ray irradiation from chromospheres, coronae, and accretion shocks associated with the central stars \citep{1997Sci...277...67K,2008A&A...492..469K,2004A&A...425..955T}. 

Unbiased single-dish radio molecular spectral line surveys have long been recognized as a powerful means to efficiently assess the molecular inventory of star-forming regions 
\citep[e.g.,][and references therein]{1984A&A...130..227J,1986ApJS...60..357B,1994ApJ...428..680B,1995ApJ...447..760V,1997ApJS..108..301S,2004PASJ...56...69K,2011A&A...532A..23C,2012ApJ...745..126W}.
In the era of the Atacama Large Millimeter Array (ALMA), such surveys loom even more important, given the need to direct high-resolution ALMA interferometric imaging studies to the brightest and/or most physically and chemically interesting molecular species. Until very recently, it would have been impractical to perform an unbiased single-dish line survey of a molecular disk, due to the general weakness of disk emission lines (a consequence of the small beam filling factors of a typical disk). With the increasing sensitivity and bandwidth of radio receivers and spectrometers, however, it is now possible to carry out such surveys --- at least for those gaseous disks that are nearest to Earth and, hence, largest in angular diameter. 

The nearby, chemically rich molecular disks orbiting TW Hya and V4046 represent two obvious subjects for such ``blind'' line surveys in the (sub)mm regime. To investigate gas physical conditions and chemistry in these two disks, and to assess the impact of stellar irradiation more specifically, we have used the 12 m  Atacama Pathfinder EXperiment (APEX\footnote{http://www.apex-telescope.org/}) telescope to perform line surveys of TW Hya and V4046 Sgr over the 0.85--1.1 mm wavelength range.

\section{Observations and Data Reduction}

\begin{figure}[!ht]
\begin{center}
\includegraphics[width=4.5in]{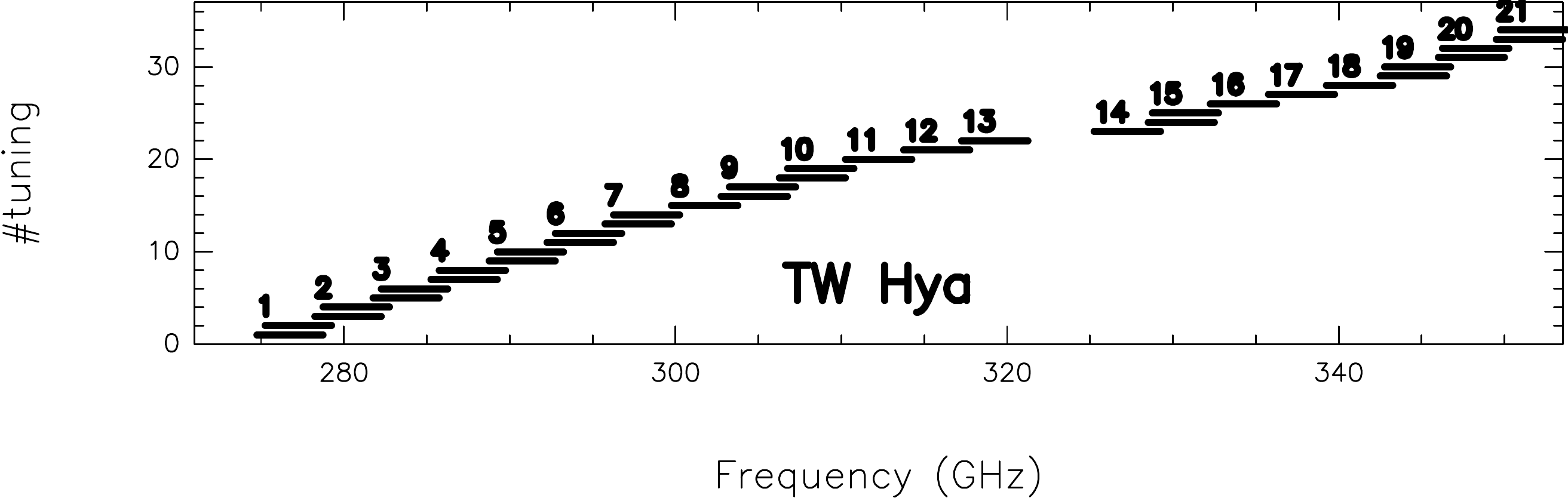}
\includegraphics[width=4.5in]{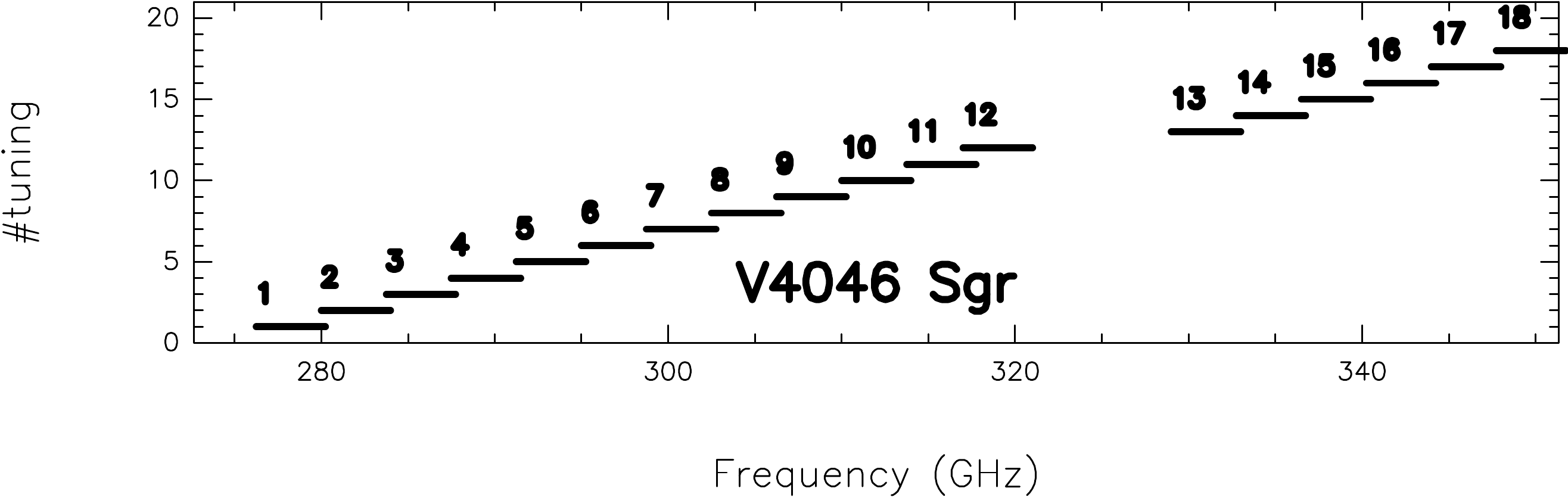}
\caption{APEX XFFTS2 spectral coverage for the TW Hya (top) and V4046 Sgr (bottom) spectral surveys. }
\label{fig:Setups}
\end{center}
\end{figure}

\begin{figure}[!hb]
\begin{center}
\includegraphics[width=3in]{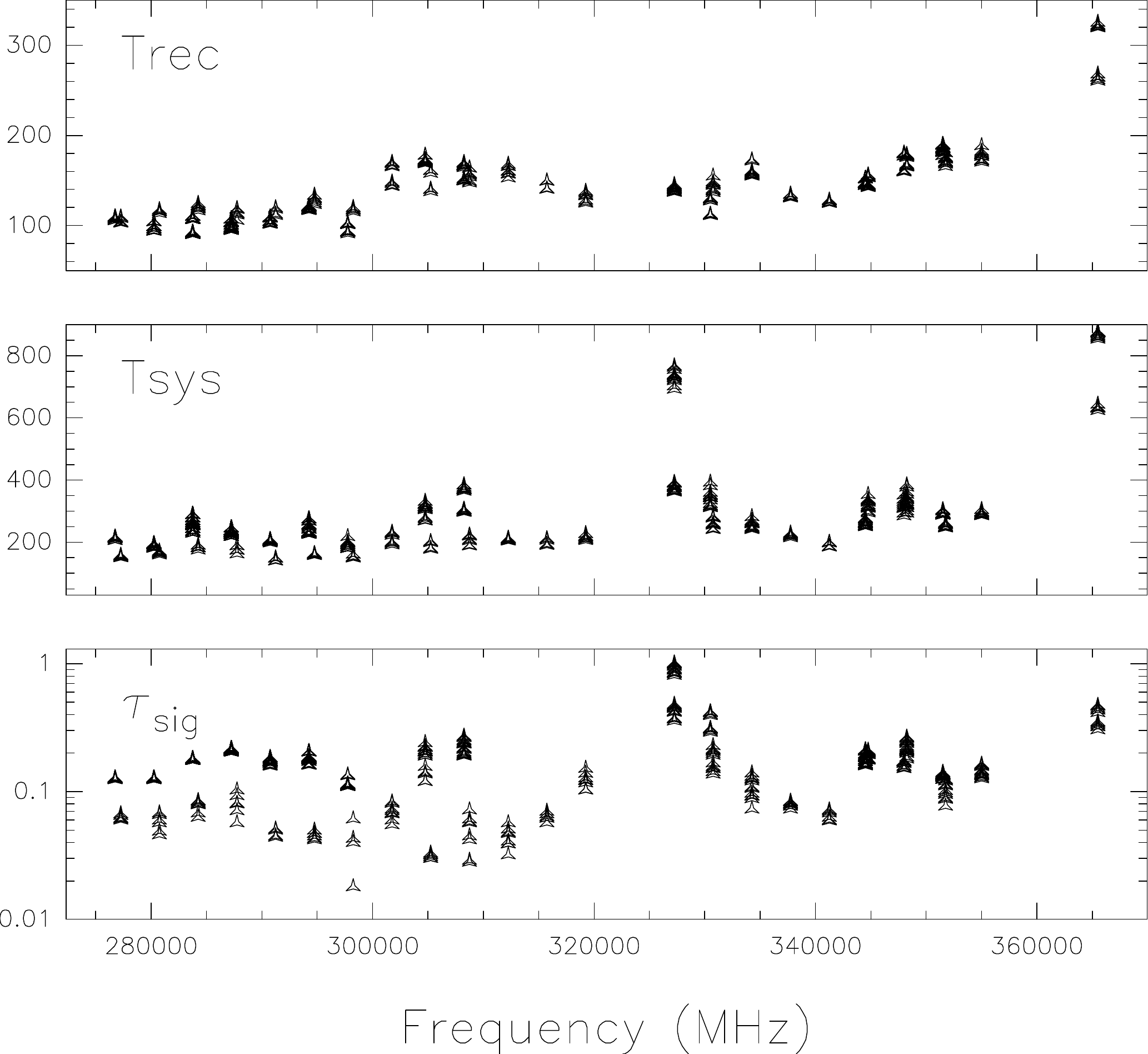}
\includegraphics[width=3in]{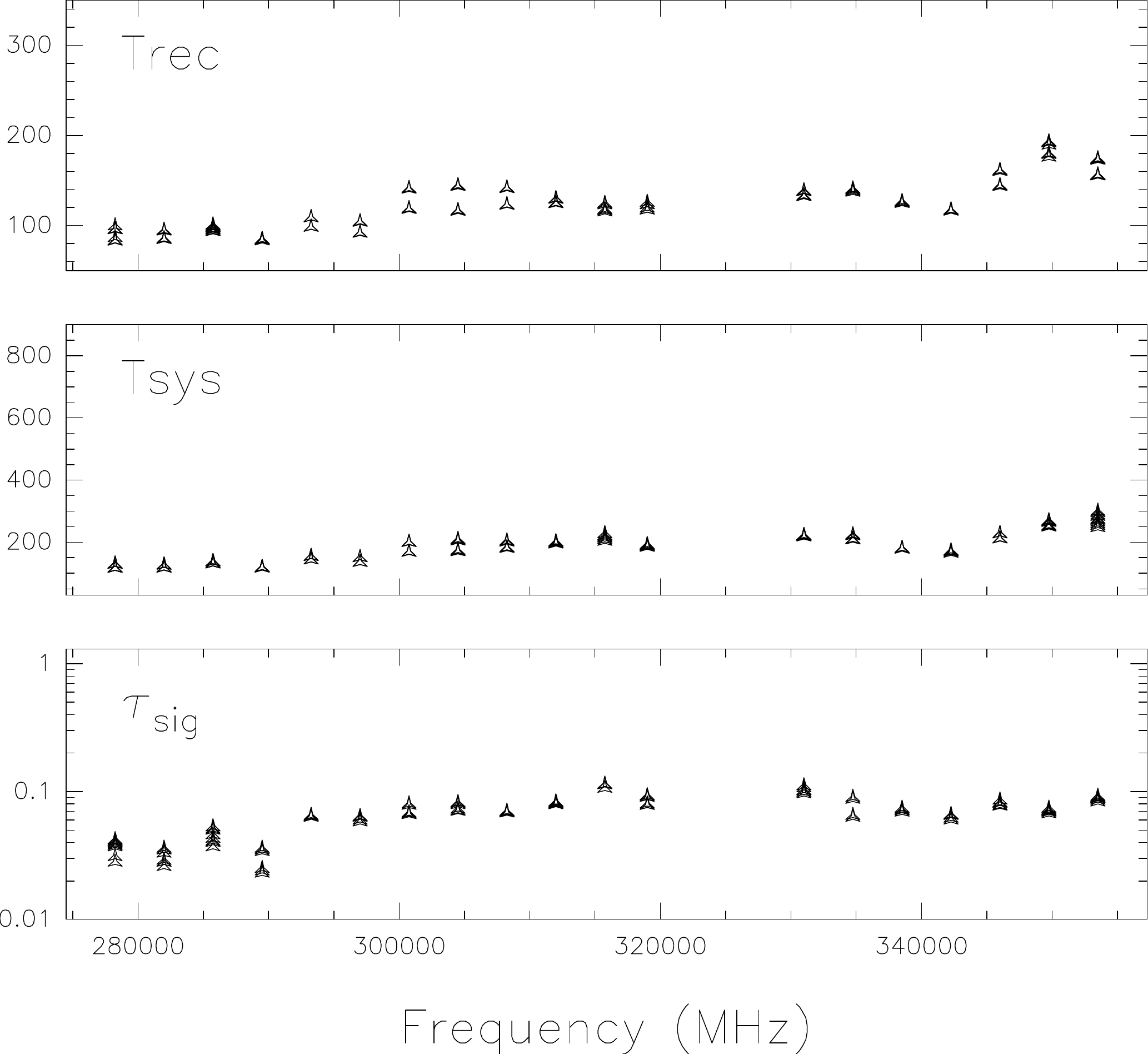}
\caption{Receiver temperatures (top), system temperatures (middle), and signal-band opacities (bottom) as functions of frequency during the TW Hya (left) and V4046 Sgr (right) spectral surveys. }
\label{fig:Conditions}
\end{center}
\end{figure}

The line surveys of TW Hya and V4046 Sgr reported here were carried out with the APEX 12 m telescope and APEX-2 receiver in 2011 December and 2012 June (TW Hya) and 2012 April and May (V4046 Sgr). The precipitable water vapor fell in the range 0.2--0.4 mm during the observations. The Fast Fourier Transform
Spectrometers \citep[XFFTS2;][]{klein2006} were used as backends, providing
4~GHz instantaneous bandwidth coverage at 77~kHz ($\approx$0.08 km s$^{-1}$) spectral
resolution.  The TW Hya and V4046 Sgr spectral surveys were conducted as sequences of 22 and 18 tuning setups, respectively, covering the 275 to 357 GHz frequency range (Fig.~\ref{fig:Setups});  the different sequences were the result of minor revisions to the observing scripts used to obtain the V4046 Sgr data. We avoided the 321 to 325 GHz interval where the atmosphere is strongly opaque. For each tuning, both polarizations were observed simultaneously, and integration proceeded until an rms channel-to-channel noise level of $\le$20 mK was achieved in 0.5 km s$^{-1}$ spectral bins. For TW Hya, observations in most tuning setups were performed twice, with a 500 MHz frequency shift between the two, before shifting the central frequency upwards by 3 GHz. Fig.~\ref{fig:Conditions} summarizes fundamental observing and instrumental conditions (i.e., receiver and system temperatures and opacities) during the spectral surveys. The approximate total on-source observing times were $\sim$1000 minutes and $\sim$310 minutes for TW Hya and V4046 Sgr, respectively; the significantly shorter time spent observing the latter source reflects the superior conditions during those observations (Fig.~\ref{fig:Conditions}).

Data reduction was performed with the CLASS software \citep{pety2005}. Reduction consisted of the standard steps, namely residual bandpass subtraction using polynomials and scan averaging with rms weighting. 
Baseline subtraction was mostly straightforward, thanks to the stability of the APEX instrumentation. Still, some so-called ``platforming'' effects were apparent; in such cases, the spectrum was split into a number of ``platforms'' from which low-order polynomials were subtracted individually. In other cases, baseline oscillations were present; these required higher-order (typically order 10) polynomial fitting, for removal. However, even such high-order fits are rather robust, given the very large number of channels (wide bandwidth) available. The averaging and stitching used a new, powerful scheme developed for CLASS spectral line data reduction\footnote{See http://www.iram-institute.org/medias/uploads/class-average.pdf}, in which each spectrum is resampled onto a common spectral axis ``on the fly'', only when needed. 

The output of the foregoing data acquisition and reduction sequences for each of the two disk sources is a single spectrum that covers the full 275--357 GHz frequency range. The results for line intensities and derived quantities described in \S 3 were obtained from these spectra, adopting an antenna temperature to flux conversion (efficiency) of 41 Jy K$^{-1}$ (corresponding to an aperture efficiency of 0.60 and a main-beam efficiency of 0.73) and a beam FWHM of $20''$; these represent mean values for APEX in the 0.8 mm window\footnote{See http://www.apex-telescope.org/instruments/.}.

\section{Results and Analysis}

\subsection{Molecular line detections}

\begin{figure}[htbp]
\begin{center}
\includegraphics[height=5in]{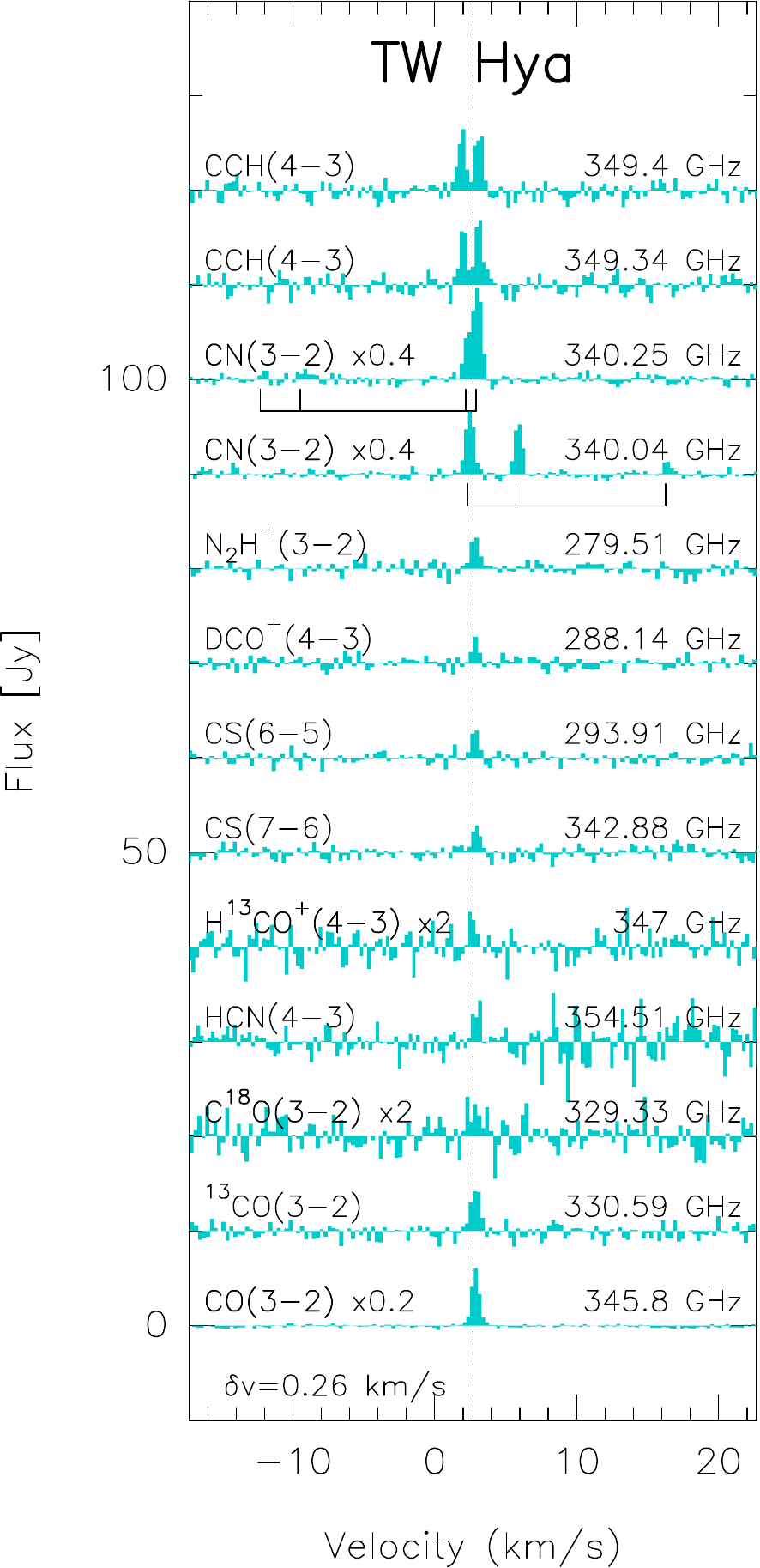}
\includegraphics[height=5in]{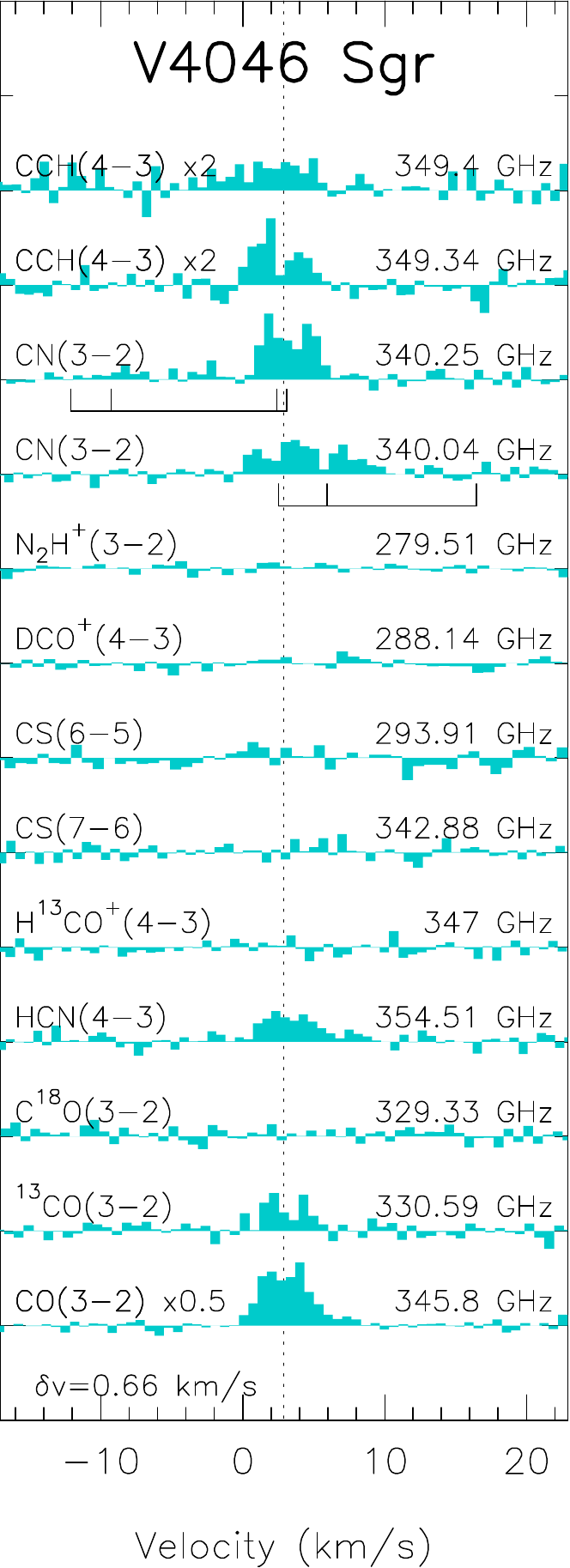}
\caption{Spectra of all molecular transitions that are well detected in the merged 0.8 mm band APEX spectrum of TW Hya (left), and the same transitions for V4046 Sgr (right). Ordinate is velocity with respect to the local standard of rest and abscissa is line flux (in Jy); for clarity, spectral regions (other than $^{12}$CO) have been shifted upwards, and some lines have been rescaled. The strongest molecular lines include C$_2$H, which is detected here for the first time in both disks (\S 3.2). }
\label{fig:AllLines}
\end{center}
\end{figure}

Using spectral line identification methods available in the WEEDS extension of the GILDAS\footnote{http://www.iram.fr/IRAMFR/GILDAS/} software tools combined with visual inspection, we have compiled lists of molecular transitions that are readily detectable and measurable in the 275--357 GHz APEX spectra of TW Hya and V4046 Sgr. A summary list of these molecular transitions is presented in Table~\ref{tbl:APEXmolecules}. Spectral regions covering all lines detected from one or both disks are displayed in Fig.~\ref{fig:AllLines}. This Figure illustrates how faint lines are more readily detectable in the APEX spectrum of TW Hya, despite the generally larger integrated line intensities measured for V4046 Sgr (Table~\ref{tbl:APEXmolecules}), thanks to the relatively narrow linewidth of TW Hya (a consequence of this disk's nearly face-on orientation; \S \ref{sec:lineIntensities}). Nevertheless, Fig~\ref{fig:AllLines} also makes clear the overall resemblance of the molecular spectra of TW Hya and V4046 Sgr in the 275--357 GHz band. Specifically, the strongest emission lines from both disks are confined to the higher-frequency (i.e., 330--355 GHz) spectral regions. The bright lines in this region include those of C$_2$H, which is detected here for first time in both the TW Hya and V4046 Sgr disks, as well as lines of species previously detected in both disks, i.e., CO, CN, and HCN \citep{1997Sci...277...67K,2008A&A...492..469K}. The CN(3--2) and C$_2$H(4--3) emission line strengths from both disks are exceeded only by that of $^{12}$CO in this frequency range (Table~\ref{tbl:APEXmolecules}; determinations of emission line intensities are described in \S 3.2). In addition, CS is detected for the first time in the TW Hya disk, in the form of emission from both the $J = 6\rightarrow5$ and $J = 7\rightarrow6$ rotational transitions.

Listings of all molecular transitions previously detected from the TW Hya and V4046 Sgr disks in the 215--365 GHz region, along with key transitions covered in the APEX line surveys reported here, are presented in Table~\ref {tbl:MolecularSpecies}. Molecular transitions measured and/or detected for the first time in our survey data are indicated in this Table. As these are the first 0.8--1.1 mm region spectral data obtained for V4046 Sgr, all lines seen in the APEX data for this disk constitute new detections. Furthermore, thanks to the broad spectral range of these line surveys, the APEX data for TW Hya and V4046 Sgr include complete coverage of all hyperfine transitions of CN and C$_2$H within a specific rotational transition ($N=3\rightarrow2$ and $N=4\rightarrow3$, respectively; see  \S \ref{sec:Hyperfine}). 

\subsection{Line intensity measurements}
\label{sec:lineIntensities}

\subsubsection{TW Hya}

For the TW Hya disk --- whose emission lines are quite narrow, due to its nearly pole-on viewing angle ($i \approx 7^\circ$; Qi et al.\ 2004) --- line intensities, LSR velocities, and widths were measured using fitting tools available in CLASS, adopting a Gaussian line profile. 
The resulting intensities are listed (in units of Jy km s$^{-1}$) in Tables~\ref{tbl:APEXmolecules} and \ref{tbl:MolecularSpecies}.  Best-fit line LSR velocities and widths (which are not listed in these tables) lie in the range 2.84--2.96 km s$^{-1}$ and 0.49--0.72 km s$^{-1}$, respectively, for all line detections with the exceptions of C$^{18}$O(3--2) and H$^{13}$CO$^+$(4--3), for which the detections are marginal (Fig~\ref{fig:AllLines}) and, hence, the best-fit line parameters are poorly determined. Given the typical fit uncertainties (0.05--0.1 km s$^{-1}$), these line-center velocities and velocity widths are consistent with each other and with those of molecular emission lines previously measured for TW Hya \citep[e.g.,][]{1997Sci...277...67K,2004A&A...425..955T}. Discrepancies between our line intensity measurements and those obtained previously are evident in a few cases (e.g., for $^{12}$CO and HCN). Such discrepancies --- which have been noted previously for certain TW Hya disk lines \citep[e.g.,][]{2004A&A...425..955T} --- are most likely due to differences in beam sizes (we have not corrected for beam filling, in converting from K km s$^{-1}$ to Jy km s$^{-1}$) as well as calibration uncertainties, which may be as large as $\sim$40\% for the weaker lines \citep[][]{2004A&A...425..955T}. 
In the case of CN and C$_2$H, we extended the line intensity modeling procedure to include a self-consistent treatment of hyperfine structure components (\S \ref{sec:Hyperfine}). Upper limits for nondetected transitions listed in Table \ref{tbl:MolecularSpecies} are 3$\sigma$ and were obtained assuming a linewidth of 0.7 km s$^{-1}$.

\subsubsection{V4046 Sgr}

The V4046 Sgr disk is viewed at intermediate inclination \citep[$i \approx 33^\circ$;][] {2010ApJ...720.1684R,2012ApJ...759..119R}; as a result, its molecular line profiles are double-peaked. Hence, to determine the intensities of lines detected in our APEX survey (Tables~\ref{tbl:APEXmolecules}), we adopt the approach of \citet{2008A&A...492..469K} to fit its line profiles. This method consists of fitting (via custom IDL codes) a parametric Keplerian line profile to determine the peak line intensity, outer disk radial velocity ($v_d$, equivalent to half of the peak-to-peak velocity width), the disk radial temperature profile power-law exponent ($q$), and outer disk density cutoff ($p_d$). Upon obtaining a good fit, one can then integrate under the model line profile to obtain the total line intensity ($I$). 

\begin{figure}[htbp]
\begin{center}
\includegraphics[width=3.5in]{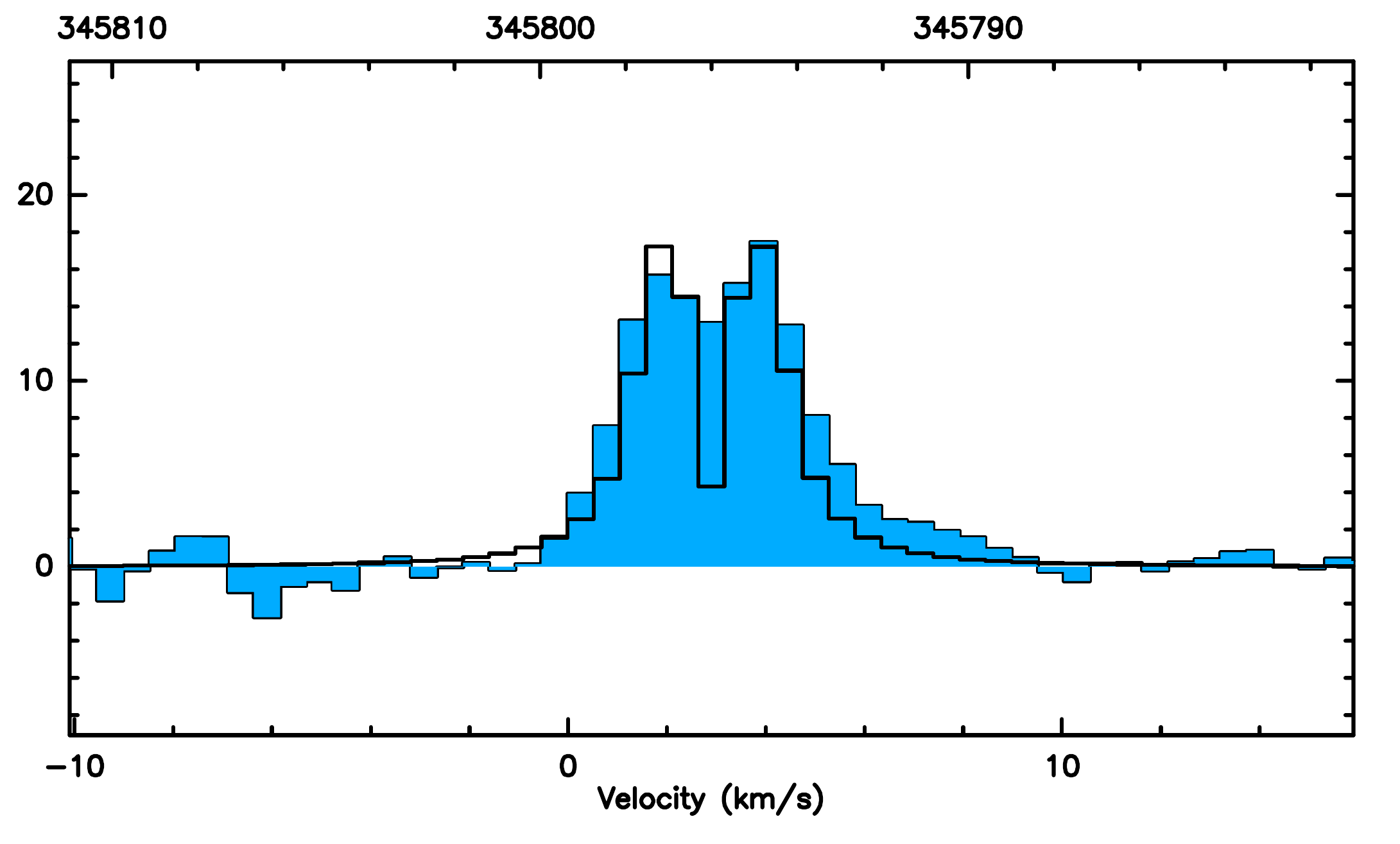}
\includegraphics[width=3.5in]{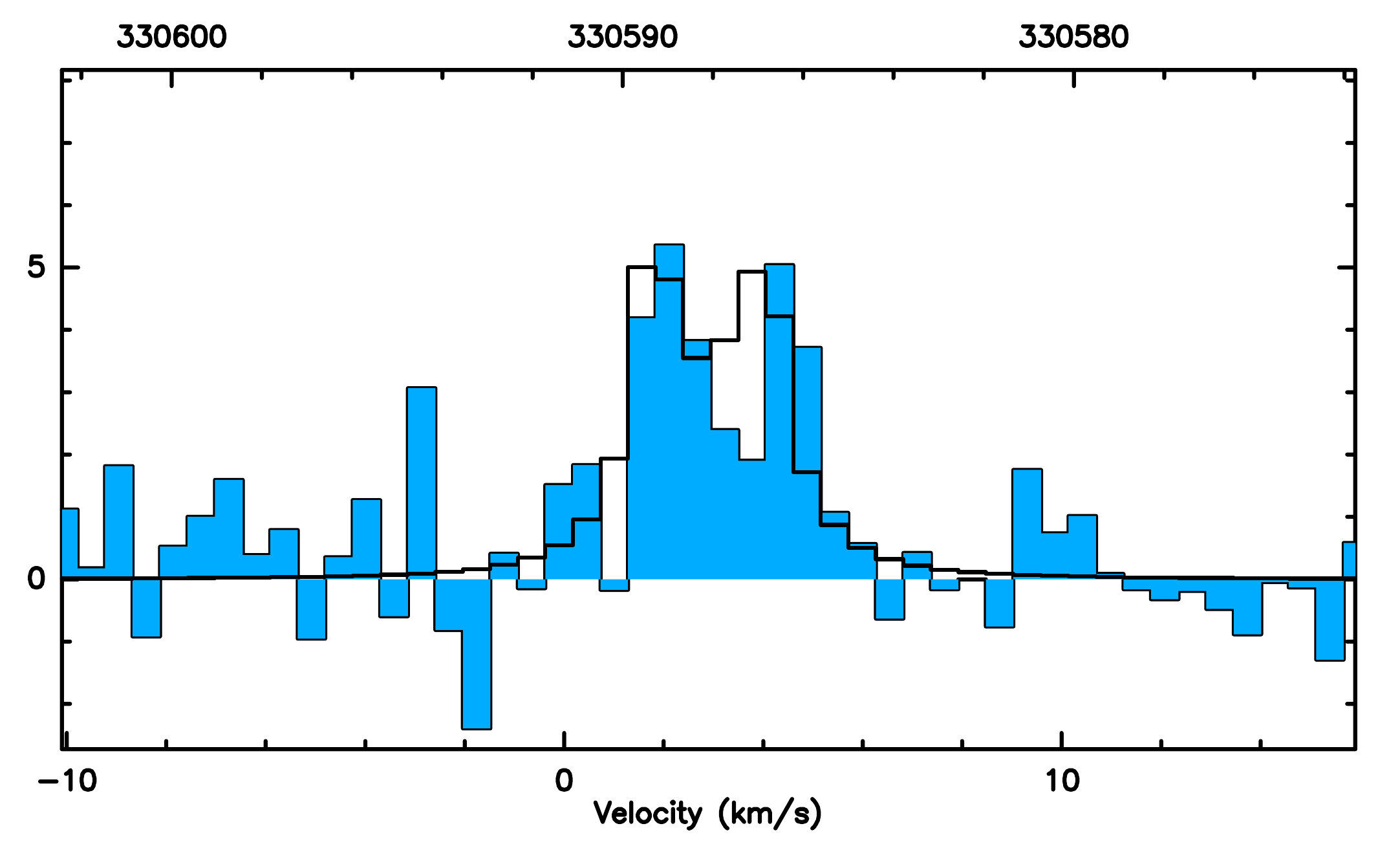}
\includegraphics[width=3.5in]{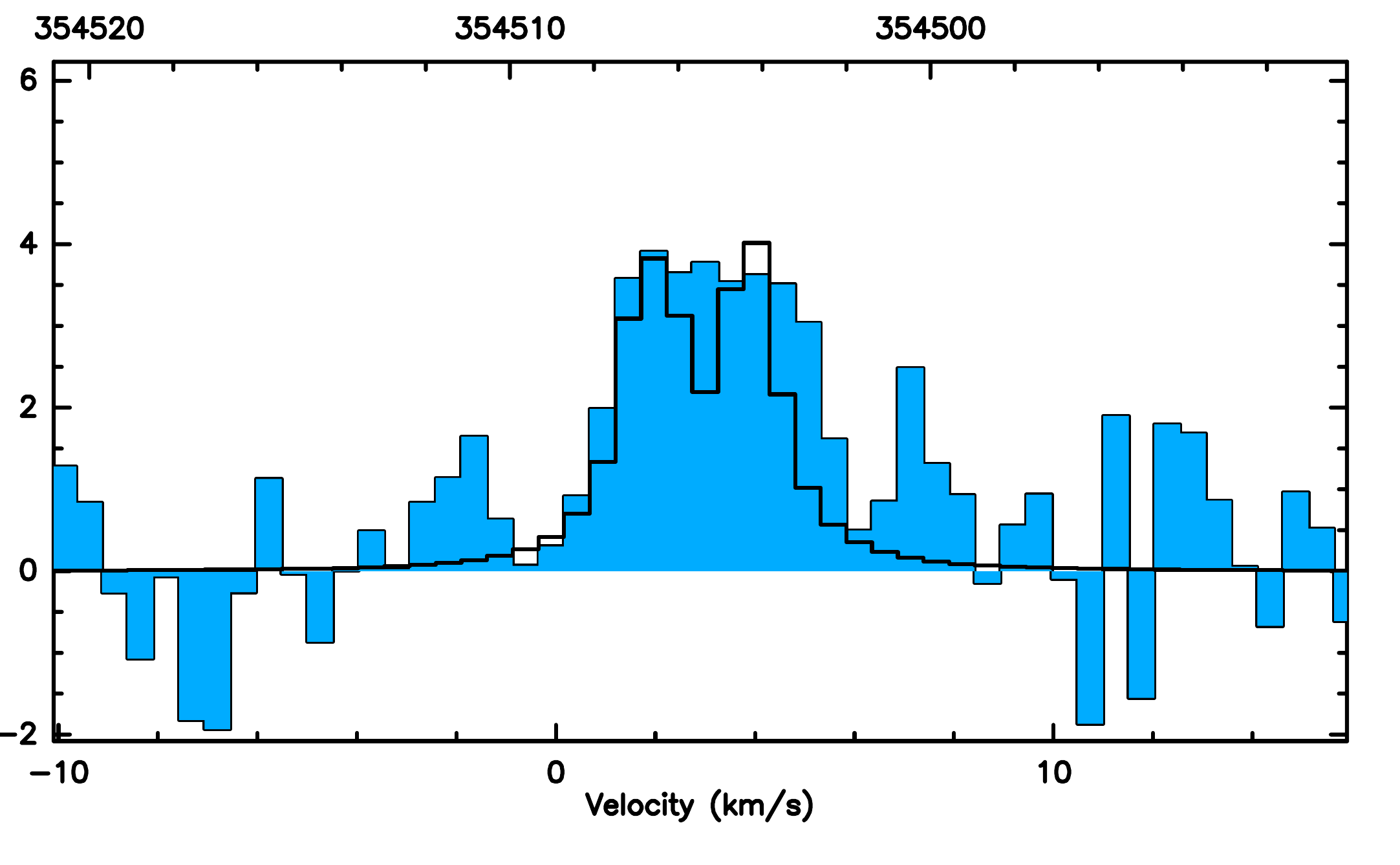}
\caption{APEX spectra of newly observed (330-354 GHz) transitions of molecular species previously detected in V4046 Sgr (blue histograms) overlaid with parametric Keplerian model line profiles (see \S 3.2.2). 
In each panel, ordinate is velocity with respect to the local standard of rest (with frequencies in MHz at the top of each panel) and abscissa is line flux in Jy. From top to bottom: CO ($J=3\rightarrow2$), $^{13}$CO ($J=3\rightarrow2$), and HCN ($J=4\rightarrow3$). }
\label{fig:V4046SgrMols}
\end{center}
\end{figure}

To determine $v_d$, we fit this parametric model to the $^{12}$CO(3--2) line, fixing $q=0.75$ and $p_d=0.25$ \citep[based on fits to lower $J$ transitions of CO measured from the V4046 Sgr disk;][]{2008A&A...492..469K}. We find a peak line intensity of 13.3$\pm$0.4 Jy and $v_d = 1.3\pm0.2$ km s$^{-1}$. The latter result is identical, to within the uncertainties, with the value determined for the $^{12}$CO(2--1) line \citep{2008A&A...492..469K}. This best-fit model is overlaid on the observed $^{12}$CO(3--2) line profile in Fig.~\ref{fig:V4046SgrMols}. Although the reduced $\chi^2$ is 1.1 for this fit, the model fails to reproduce the (shallow) dip between the peaks in the $^{12}$CO line profile, and underestimates the emission in the profile's (broad) red wing. These deficiencies are likely the result of the oversimplifications inherent in the parametric Keplerian model profile. As our purpose here is to obtain line intensity measurements (as opposed to physical inferences as to disk structure), however, the results described in subsequent sections are unaffected.
To obtain intensity measurements of the other (much noisier) lines, including $^{13}$CO(3--2), we fixed the model parameters to the foregoing values of $v_d$, $q$, and $p_d$, such that the only free parameter of the fits was the peak line intensity. The resulting fits to the $^{13}$CO(3--2) and HCN(4--3) line profiles --- two species that, like $^{12}$CO, were previously detected in the V4046 Sgr disk --- are also shown in Fig.~\ref{fig:V4046SgrMols}. The best-fit peak line intensities for $^{13}$CO(3--2) and HCN(4--3) are 4.0$\pm$0.3 Jy and 3.0$\pm$0.2 Jy, respectively, with reduced $\chi^2$ values of 1.1 and 0.8. Upper limits for nondetected transitions  listed in Tables \ref{tbl:APEXmolecules} and \ref{tbl:MolecularSpecies} are 3$\sigma$ and were obtained assuming the same value of $v_d$ as found for $^{12}$CO(3-2), i.e., $v_d = 1.3$ km s$^{-1}$.




\subsection{Pure rotational transitions: column density estimates}
\label{sec:colDens}

We have used the results for integrated line intensities for the (pure rotational) transitions of $^{13}$CO, HCN, and CS listed in Table~\ref{tbl:APEXmolecules} (as well as the molecular line reference data listed in that Table) to estimate source-averaged column densities for these molecules, using the methods described in \citet{1999ApJ...517..209G}. Briefly, this involves calculating the upper level column density $N_J$ for the transition $J\rightarrow (J-1)$ via
\begin{equation}
N_J = \frac{8\pi k \nu^2 W}{hc^3 A_{J,J-1}}  (\frac{\tau}{1-e^{-\tau}}) \fbeam,
\end{equation}
where $\nu$ is the frequency of the transition, $W$ is the integrated line intensity (in K cm s$^{-1}$), $\tau$ is the optical depth in the transition, $\fbeam = \frac{\Omega_{mb}+\Omega_s}{\Omega_s}$ is the main-beam dilution correction factor
(with $\Omega_{mb}$ and $\Omega_s$ the main-beam and source solid
angles, respectively), and $A_{J,J-1}$ is the Einstein coefficient for the transition, given by
\begin{equation}
A_{J,J-1} = \frac{64 \pi^4 \nu^3 \mu^2}{3 h c^3}\frac{J}{2J+1},
\end{equation}
with $\mu$ the permanent electric dipole moment of the molecule. The upper level column density can then be converted to total column density for a given molecular species via
\begin{equation}
N = N_J \; \frac{Q(T_{ex})}{g_J} \; e^{E_J/kT_{ex}},
\end{equation}
where the temperature-dependent rotational partition function is obtained from
\begin{equation}
Q(T_{ex}) \approx \frac{1}{3} + \frac{kT_{ex}}{hB_{rot}}, 
\end{equation}
with $B_{rot}$ the molecular rotational constant. Eq.\ 4 is appropriate for linear (and approximately linear) molecules, i.e., for molecules with only one (or only one dominant) moment of inertia.

We used the foregoing methods to obtain source-averaged column densities $N$ for a range of $T_{ex}$ of 5--37.5 K. The upper end of this $T_{ex}$ range is perhaps $\sim$30\% larger than the CO line excitation temperatures previously estimated for both disks \citep[e.g.,][]{2004ApJ...616L..11Q,2013ApJ...775..136R}, while the lower end roughly corresponds to the excitation temperature of C$_2$H as deduced from our analysis of C$_2$H hyperfine structure emission from TW Hya (\S \ref{sec:Hyperfine}).  We assume optically thin emission, and adopt identical beam dilution factors of $\fbeam=0.25$ for both disks and for all molecules in each disk. The latter is obtained by assuming that the characteristic emitting region radii $R$ for all five molecules are the same as that of $^{12}$CO, i.e., $R \approx 5''$ for both TW Hya and V4046 Sgr \citep{2012ApJ...757..129R,2013ApJ...775..136R}. Since less abundant molecules likely have smaller effective beam filling factors than $^{12}$CO \citep[see, e.g., Fig.\ 6 in][]{2013ApJ...775..136R}, the source-averaged column densities of these species may be somewhat underestimated.

Results for $N$ for $^{13}$CO, HCN, and CS are listed in Table~\ref{tbl:ColumnDensities}. The $^{13}$CO  column density we find for the TW Hya disk for the upper range of $T_{ex}$, $N$($^{13}$CO)$\sim2\times10^{14}$ cm$^{-2}$, is a factor $\sim$2.5 smaller than the $^{13}$CO  column density previously determined by \citet{2004A&A...425..955T} under the assumption $T_{ex} = 25$ K. This difference in $N$($^{13}$CO) values is mostly accounted for by the difference in assumed source radii \citep[][adopted a source radius of $\sim$3$''$]{2004A&A...425..955T}. The (disk-averaged) $^{13}$CO column density we obtain for TW Hya assuming $T_{ex} = 18.75$ K and a beam filling factor $\fbeam=0.25$ is furthermore consistent with the radially-dependent $N$($^{13}$CO) values at large ($R\sim100$ AU) disk radii determined by \citet{2013Sci...341..630Q} from fits of a detailed disk model to interferometric $^{13}$CO data. Our HCN column density in this same $T_{ex}$ range, $N$(HCN)$\sim5\times10^{11}$ cm$^{-2}$, is more than an order of magnitude smaller than the value of $N$(HCN) reported by \citet{2004A&A...425..955T}. This discrepancy is evidently due to their assuming optically thick HCN emission, in addition to a smaller source radius.  

The estimated $^{13}$CO column densities for V4046 Sgr are somewhat larger than those estimated for TW Hya (Table~\ref{tbl:ColumnDensities}). This result --- combined with the fact that the V4046 Sgr disk is also almost a factor 2 larger than that of TW Hya in linear radius --- is consistent with various other indications that the V4046 Sgr molecular disk is more massive than that of TW Hya \citep{2008A&A...492..469K,2013ApJ...775..136R,2013Natur.493..644B}, despite the (factor $\sim$2) greater age of the former system.

\subsection{CN and C$_2$H hyperfine component analysis: optical depths, excitation temperatures, and column densities}
\label{sec:Hyperfine}

\subsubsection{TW Hya}

\begin{figure}[htbp]
\begin{center}
  \includegraphics[width=0.9\hsize]{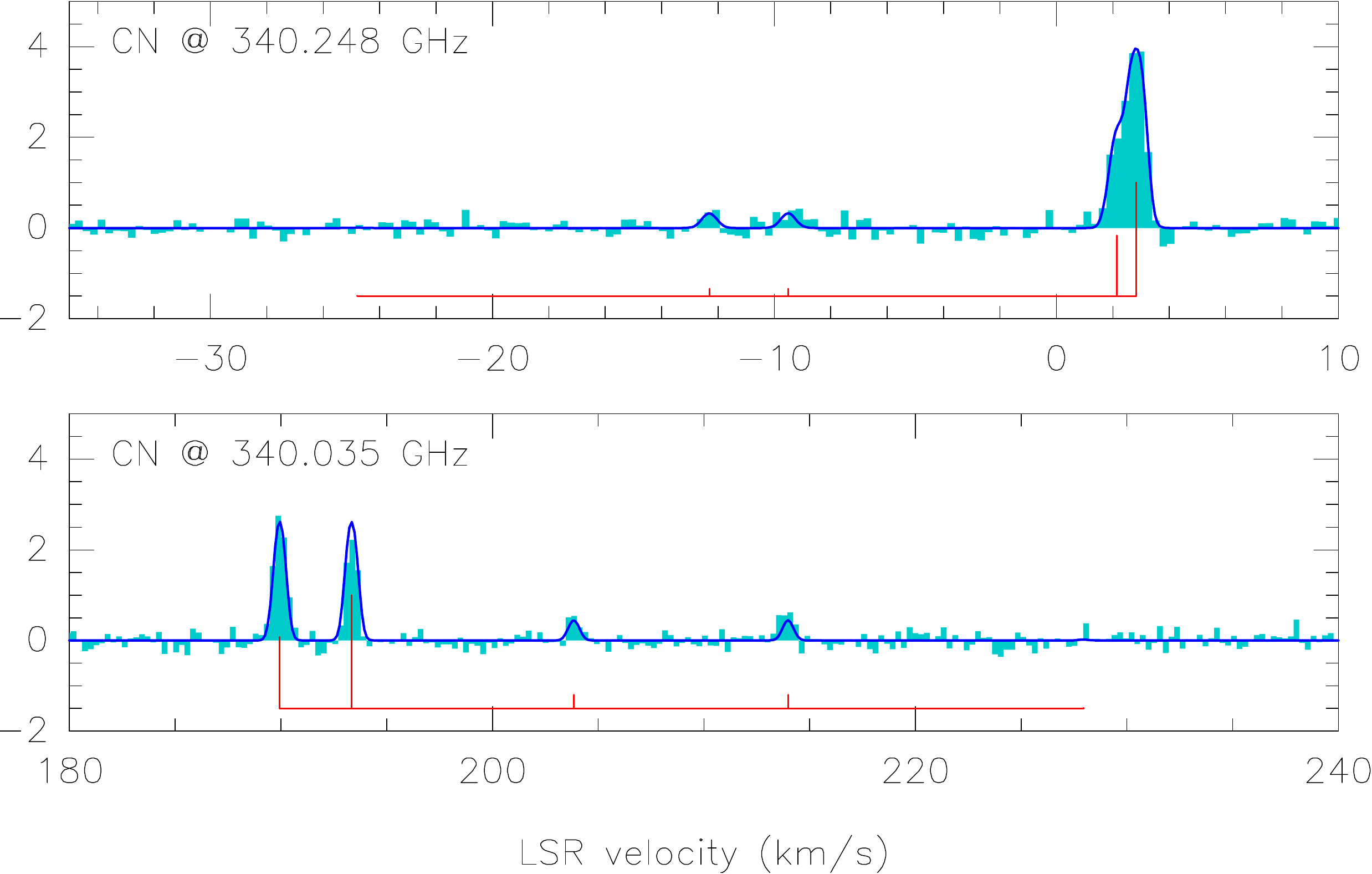}\bigskip\\
  \includegraphics[width=0.9\hsize]{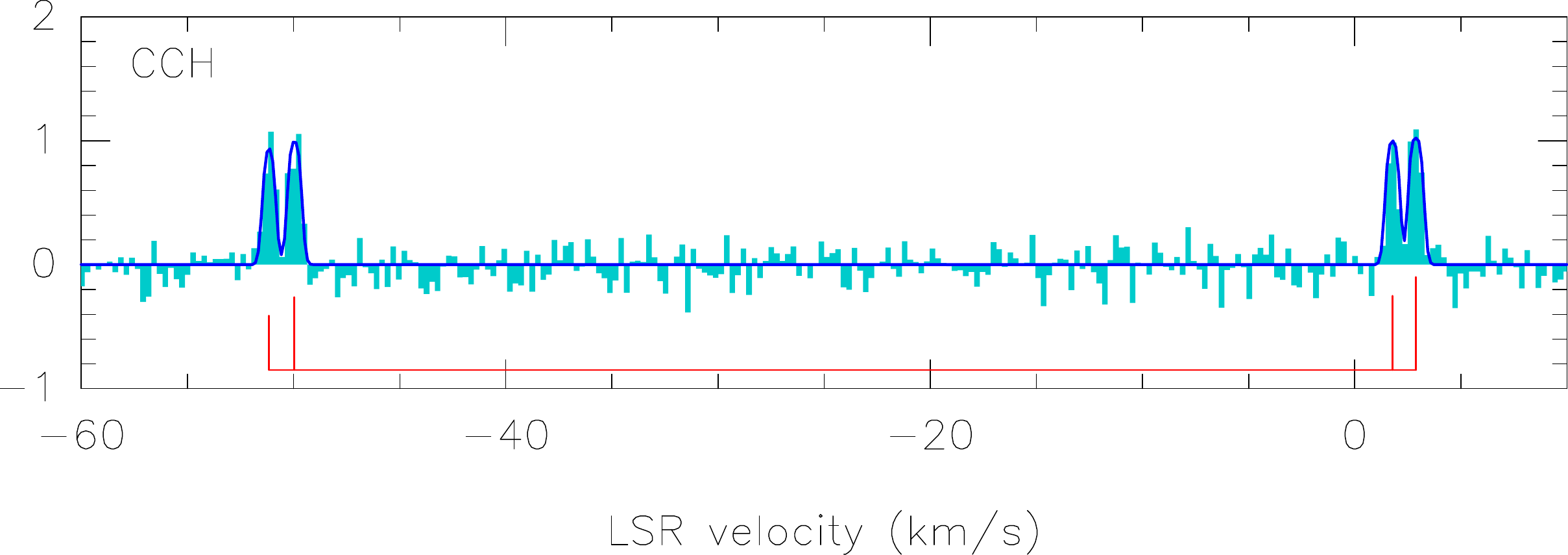}
  \caption{Full spectra of hyperfine components of CN (top and middle panels) and C$_2$H (bottom panel) observed toward TW Hya. The red lines indicate the positions and relative intensities of the hyperfine components (see Table~\ref{tbl:CNandCCHresults}). The spectra are overlaid with the best-fit models obtained via the method described in \S \ref{sec:Hyperfine}. The intensity scale (in K) has been corrected for beam efficiency and beam dilution.}
\label{fig:TWHyaCNCCH}
\end{center}
\end{figure}

The rotational state energy levels of the molecules CN and C$_2$H exhibit fine structure and hyperfine structure splitting (HFS) due to interactions between molecular rotational angular momentum with the electron and nuclear spins; these interactions result in a sequence of closely-spaced lines for each rotational state transition in the mm and submm \cite[Table~\ref{tbl:CNandCCHresults}; see also, e.g.,][]{1982ApJ...254...94Z}. We performed measurement and analysis of the TW Hya CN $N=3\rightarrow2$ and C$_2$H $N=4\rightarrow3$ spectra --- each of which is displayed, in its entirety, in Fig.~\ref{fig:TWHyaCNCCH} --- using the classical HFS fitting method implemented in the CLASS software. This code makes the simplifying assumption that all hyperfine transitions within a given rotational transition share the same excitation temperature. For optically thin emission, the measured line ratios thus depend solely on the relative line strengths, which scale as the product $g_u A_{ul}$ (where $g_u$ and $A_{ul}$ are the degeneracy of the upper hyperfine level and Einstein coefficient of the hyperfine transition, respectively). In addition to obtaining hyperfine component intensity measurements, simultaneously fitting all hyperfine components results in determinations of the opacity and excitation temperature of the rotational transition. Fig.~\ref{fig:TWHyaCNCCH} and Table~\ref{tbl:CNandCCHresults} summarize the results of this analysis for the APEX measurements of CN and C$_2$H line emission from TW Hya. The derived values of $\tau$ and $T_{\rm ex}$ are listed in Table~\ref{tab:hfs}. 

Determination of column densities for hyperfine structures requires special care \citep[see, e.g.,][]{2004A&A...425..955T}. In particular, the total partition function $Q_{\rm tot}$ must include the nuclear-spin statistics. In the case of the CN radical, this results in a factor 6 increase in $Q_{\rm tot}$ due to the $I=1$ nuclear spin of the N atom and the $S=1/2$ spin of the unpaired electron \citep{skatrud1983}. The C$_2$H radical also has a $^2\Sigma^+$ ground electronic state, hence presenting a spin-rotation splitting, which further couples with the nuclear spin of the H atom \citep{muller2000}, such that the total partition function is 4 times the rotational partition function. The column density is then computed as 
\begin{equation}
  N_{\rm tot} = \frac{8\pi\nu^3}{c^3} \,
  \frac{e^{E_l/kT_{\rm ex}}}{1 - e^{-\tul/T_{\rm ex}}}\,
  \int \tau_{ul} dv
  \times
  \frac{Q_{\rm tot}(T_{\rm ex})}{\aul\gup}
  \label{eq:ntot-hf}
\end{equation}
where all parameters refer to the particular hyperfine transition of interest. The
beam-dilution correction (see Eq. 1) is applied to the spectrum before the HFS
fitting procedure is performed. As for the column density estimates for $^{13}$CO, HCN, and CS (\S 3.3), we adopt a beam dilution correction factor $\fbeam=0.25$, although the results are relatively insensitive to the assumed source solid angle; e.g., assuming a C$_2$H source radius that is smaller by a factor of 2 results in a factor $\sim$2 increase in $T_{ex}$ (for CN, the value of $T_{ex}$ is even less dependent on source solid angle). Following this procedure, we obtain estimates of $N_{\rm tot}$(CN) $\sim10^{14}$ cm$^{-2}$ and $N_{\rm tot}$(C$_2$H) $\sim5\times10^{15}$ cm$^{-2}$ for the TW Hya disk (Table~\ref{tab:hfs}). The former estimate is somewhat larger than that obtained by \citet{2004A&A...425..955T}, who based their results on a single CN hyperfine transition and assumed $T_{ex}=25$ K (see \S 4).

\subsubsection{V4046 Sgr}

\begin{figure}[htbp]
\begin{center}
\includegraphics[width=4in]{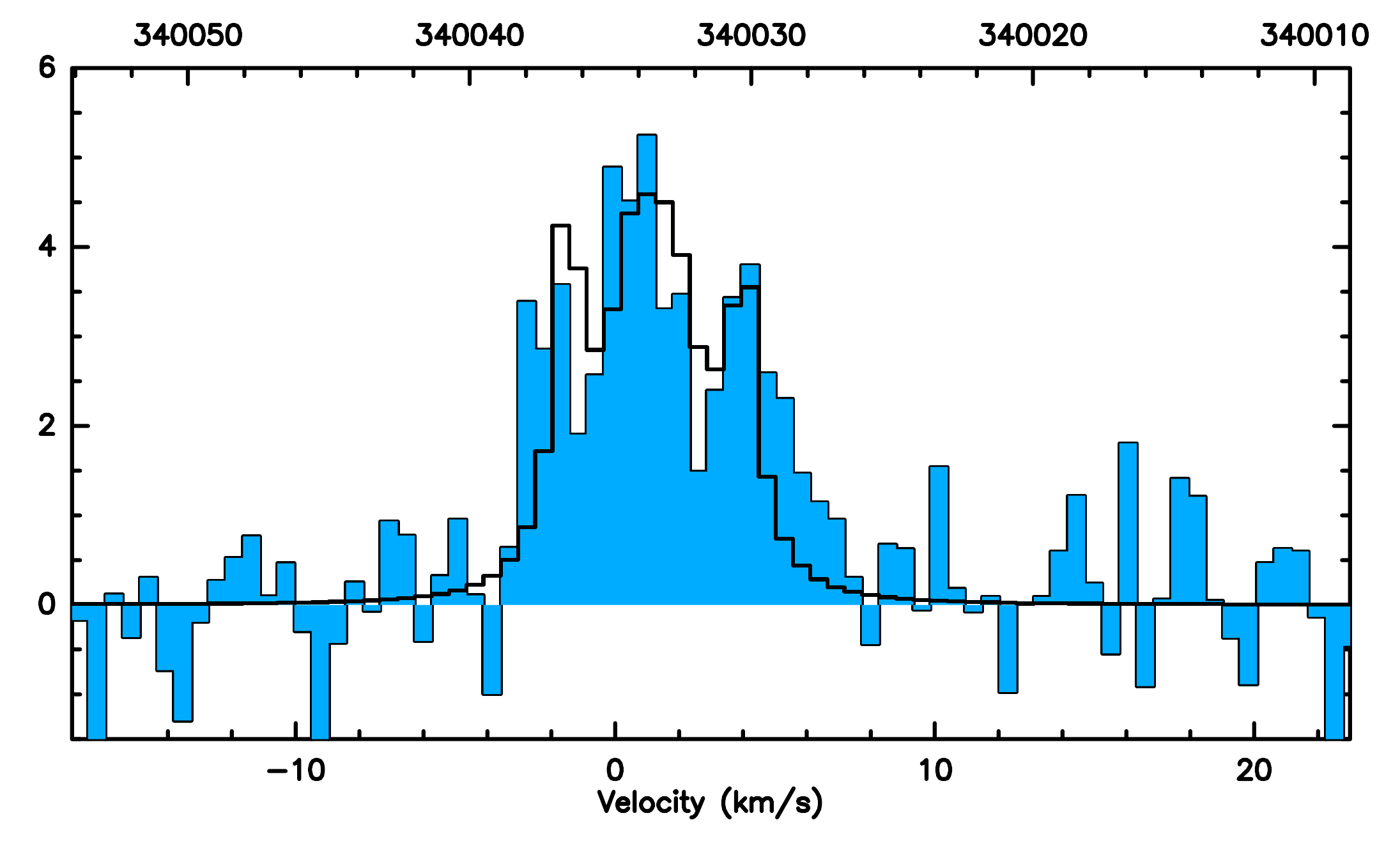}
\caption{Spectrum of  340.04 GHz CN ($N=3\rightarrow2$) 
emission from the V4046 Sgr disk, with best-fit multiple-component parametric Keplerian line profile overlaid (see \S 3.4.2). Ordinate is
  velocity with respect to the local standard of rest (with corresponding frequencies in MHz labeled across the top of the panel) and abscissa is line flux (in Jy).}
\label{fig:V4046SgrCNCCH}
\end{center}
\end{figure}

In the spectrum of V4046 Sgr, only the brighter components of the CN and C$_2$H hyperfine structure lines are detected (the detected components are displayed in the top panels of Fig.~\ref{fig:AllLines}).  Furthermore --- as is evident from the comparison with TW Hya in Fig.~\ref{fig:AllLines} --- these bright hyperfine components are blended, as a consequence of the V4046 Sgr disk's broadened, double-peaked line profiles. We attempted to estimate intensities for the combined fluxes of the brightest hyperfine components by fitting a two-component Keplerian line profile model, in which the free parameters are the peak temperature of the brighter component and the component line ratio. The resulting total line intensities are listed in Table~\ref{tbl:CNandCCHresults}, and the two-component fit to the 340.04 GHz hyperfine complex of CN is displayed in Fig.~\ref{fig:V4046SgrCNCCH}. Even in this case, where the line blending is apparent in the observed profile, the data are sufficiently noisy that the line ratio is poorly constrained. Hence, in contrast to TW Hya, we cannot confidently determine the range of $\tau$ and $T_{ex}$ characterizing the CN and C$_2$H emission from V4046 Sgr. However, we note that the results for the blended line intensities in these brighter hyperfine components suggest that the CN and C$_2$H emission is optically thin and thick, respectively. That is, whereas the CN line ratios are consistent with the theoretically predicted ratios (Table~\ref{tbl:CNandCCHresults}), the ratio of the two main C$_2$H component line complexes is consistent with nearly equal line intensities among the hyperfine components. The latter result suggests that we can estimate the excitation temperature of C$_2$H from its peak measured main-beam antenna temperature. Adopting the same correction for beam dilution as applied in the case of TW Hya (i.e., $\fbeam=0.25$), and accounting for the cosmic microwave background, we estimate $T_{ex} \sim$ 4 K, i.e., very similar to the (low) $T_{ex}$ value determined for TW Hya via analysis of its C$_2$H hyperfine emission (\S 3.4.1; Table~\ref{tab:hfs}).

As we have only limited information concerning $\tau$ and $T_{ex}$ for CN and C$_2$H in the case of V4046 Sgr, we can only obtain order-of-magnitude estimates for the column densities for these species. Scaling the results for TW Hya (Table~\ref{tab:hfs}) according to the relative integrated line intensities, under the assumption that $\tau$ and $T_{ex}$ are similar for the two disks, we crudely estimate $N_{\rm tot}$(CN) $\sim10^{14}$ cm$^{-2}$ and $N_{\rm tot}$(C$_2$H) $\sim10^{16}$ cm$^{-2}$ for V4046 Sgr. The former is likely an upper limit since, in the case of V4046 Sgr, the CN lines are evidently not optically thick.

\subsection{Line ratios of isotopologues of CO: opacity diagnostics}

Based on the sharp departures of measured $^{13}$CO:$^{12}$CO line ratios from the canonical $^{13}$C:$^{12}$C isotopic ratio determined for the solar neighborhood, \citet{1997Sci...277...67K} and \citet{2004A&A...425..955T} had previously estimated $^{12}$CO(3--2) emission line optical depths in the range $\tau_{^{12}CO}\sim10$ for TW Hya; \citet{2008A&A...492..469K} similarly estimated $\tau_{^{12}CO}\sim30$ for V4046 Sgr, based on measurements of the $J= 2 \rightarrow 1$ transitions of $^{13}$CO and $^{12}$CO. However, these estimates rely on the assumption that emission from $^{13}$CO is optically thin. The measurements of C$^{18}$O(2--1) and (3--2) emission from the TW Hya disk in \citet{2013Sci...341..630Q} and Table~\ref{tbl:APEXmolecules} provide a test of this assumption since, under the assumption of identical $^{13}$CO and C$^{18}$O excitation temperatures,  
\begin{equation}
R = \frac{1-\exp{(-\tau_{^{13}CO})}}{1-\exp{(-\tau_{^{13}CO}/X)}},
\end{equation}
where $R$ is the measured $^{13}$CO:C$^{18}$O line ratio and $X$ is the $^{13}$CO:C$^{18}$O abundance ratio. The latter quantity can be obtained from the ratio of isotopic ratios, i.e., $^{13}$C:$^{12}$C/$^{18}$O:$^{16}$O. We adopt $X=7$, based on the values $^{12}$C:$^{13}$C $=68$ and $^{16}$O:$^{18}$O $=480$ \citep[as determined from measurements of CN and CO lines for the local interstellar medium and Sun, respectively;][]{2005ApJ...634.1126M,2006A&A...456..675S}. Hence, in the optically thin limit ($\tau_{^{13}CO}<<1$), we expect $R \approx 7$. The measured ratios for the $J=2\rightarrow 1$ and $J=3\rightarrow 2$ transitions for TW Hya ($R\approx 4.0$ and $R\approx 2.0$, respectively; Table~\ref{tbl:APEXmolecules}), therefore imply the emission in the $^{13}$CO(2--1) and (3--2) lines is optically thick, with implied optical depths of $\tau_{^{13}CO} \sim 1.5$ and $\tau_{^{13}CO} \sim 5$, respectively. 

Given the foregoing assumption for the isotopic abundances of carbon, we further expect $\tau_{^{12}CO}\approx68 \tau_{^{13}CO}$. Hence, it appears that previous single-dish-based determinations of $\tau_{^{12}CO}$ for TW Hya are severe underestimates. If both $^{12}$CO(3--2) and $^{13}$CO(3--2) are indeed optically thick, then the mere factor $\sim$7 difference in their integrated line intensities (Table~\ref{tbl:APEXmolecules}) could reflect the fact that the (optically thicker) $^{12}$CO emission arises from higher, warmer disk layers than the $^{13}$CO. It is also likely, however, that the implicit assumption of uniform $^{13}$CO optical depth over the disk surface is an oversimplification. We further note that our upper limit on C$^{18}$O(3--2) emission from V4046 Sgr suggests $^{13}$CO:C$^{18}$O $> 2$, which would imply $\tau_{^{13}CO} < 5$ for V4046 Sgr in the $J=3\rightarrow 2$ transition --- a surprising result, given the previous indications that $^{12}$CO emission from TW Hya is, if anything, more optically thick than that from V4046 Sgr. The contrast in CO optical depths between the TW Hya and V4046 Sgr disks may be related to different outer disk density structures and/or outer radii, as the TW Hya molecular disk is somewhat smaller than the disk orbiting V4046 Sgr \citep[respective radii $\sim$200 AU and $\sim$350 AU;][]{2012ApJ...744..162A,2013ApJ...775..136R}. The two disks also have sharply contrasting disk structures on radial scales $<50$ AU (i.e., $<$1$''$), but such small-scale differences are unimportant for our single-dish ($\sim$20$''$ beam diameter) CO measurements. Clearly, sensitive interferometric observations of $^{13}$CO and C$^{18}$O toward both disks are warranted \citep[see also discussion in][]{2013Sci...341..630Q}.

\section{Discussion}

In Table~\ref{tbl:ColDensComp}, we list fractional molecular abundances relative to $^{13}$CO, $N$(X)/$N$($^{13}$CO), obtained from the determinations of source-averaged molecular column densities for TW Hya and V4046 Sgr described in \S 3.4 and listed in Tables~\ref{tbl:ColumnDensities} and \ref{tab:hfs}. These values of $N$(X)/$N$($^{13}$CO) were obtained by adopting $T_{ex} = 18.75$ K for $^{13}$CO \citep[based on recent modeling of the TW Hya and V4046 Sgr disks;][]{2013ApJ...775..136R,2013Natur.493..644B} as well as for HCN and CS, and $T_{ex}= 10$ K and 5 K for CN and C$_2$H, respectively (Table~\ref{tab:hfs}). 
Table~\ref{tbl:ColDensComp} also lists values of $N$(X)/$N$($^{13}$CO) based on results previously obtained for the disks orbiting the T Tauri stars LkCa 15 and DM Tau as well as for the deeply embedded (candidate ``Class 0''), low-mass protostars R~CrA~IRS~7B and IRAS 16293--2422. To our knowledge, these four objects are the only other low-mass, pre-main sequence stars and protostars for which molecular line surveys encompassing all five of the molecules in Tables~\ref{tbl:ColumnDensities} and \ref{tab:hfs} have been reported in the literature. Both of the protostars are classified as ``hot corinos,'' i.e., protostellar envelopes that are particularly chemically active due to strong heating from accretion processes and/or external radiation fields \citep[e.g.,][]{2007A&A...463..601B,2012ApJ...745..126W}. Table~\ref{tbl:ColDensComp} also includes fractional molecular abundances measured for the planetary nebula NGC 7027, which features strong stellar UV and nebular X-ray radiation fields \citep[see, e.g.,][]{2012AJ....144...58K}.  

The comparison of $N$(X)/$N$($^{13}$CO) values in Table~\ref{tbl:ColDensComp} 
indicates that the fractional CN and C$_2$H abundances in the evolved disks orbiting the ``old'' (age $\sim$8--20) Myr-old T Tauri stars TW Hya and V4046 Sgr are strongly (factors $\sim$30--1000) enhanced over those characteristic of the other chemically rich object classes represented in the Table; C$_2$H appears to be even more abundant in the TW Hya and V4046 Sgr disks than in the somewhat younger (age 3--5 Myr) T Tauri star/disk systems LkCa~15 and DM Tau. We caution that the apparent sharp contrast in the ratios $N$(CN)/$N$($^{13}$CO) and (especially) $N$(C$_2$H)/$N$($^{13}$CO) that are listed in Table~\ref{tbl:ColDensComp} may be due, at  least in part, to different assumptions regarding the beam filling factors and optical depths of emission from the various molecular species in these sources, as well as to the variety of methods used to deduce $T_{ex}$ for purposes of calculating the values of $N$. Nevertheless, this comparison between disks, ``hot corinos,'' and UV/X-ray-luminous planetary nebula supports previous suggestions \citep{1997Sci...277...67K,2008A&A...492..469K,1997A&A...317L..55D,2004A&A...425..955T,2010ApJ...714.1511H} that high-energy (FUV and/or X-ray) radiation from the vicinity of the central stars is enhancing the abundances of CN and C$_2$H in T Tauri disks. Similar inferences concerning the likely influence of X-rays on disk chemistry were drawn by   \citet{2008A&A...492..469K}, \citet{2004A&A...425..955T}, and \citet{2011A&A...536A..80S} on the basis of enhanced HCN and HCO$^+$ line fluxes from T Tauri disks.  Our tentative determinations of low $T_{ex}$ ($\sim$5--10 K) for CN in TW Hya and for C$_2$H in both TW Hya  and V4046 Sgr (\S 3.4) --- which are consistent with inferences concerning $T_{ex}$ for these two species in the DM Tau and LkCa~15 disks \citep{2010ApJ...714.1511H,2012A&A...537A..60C} --- may indicate that the mm-wave emission from CN and C$_2$H arises from regions that are rather deep within the disks. If so, it would then stand to reason that the abundances of C$_2$H and, perhaps, CN are being enhanced by X-rays rather than by FUV photons, given that X-rays can penetrate to much larger column densities within the disk (e.g., Skinner \& Guedel 2013; Cleeves, Adams, \& Bergin 2013).

Indeed, all four stars considered here are relatively luminous X-ray sources --- with $L_X \approx 10^{30}$ erg s$^{-1}$ in each case --- based on published X-ray studies for TW Hya \citep{2002ApJ...567..434K}, V4046 Sgr \citep{2006A&A...459L..29G}, and LkCa~15 \citep{2013ApJ...765....3S}, and archival X-ray data for DM Tau (which was detected serendipitously by Chandra). Furthermore, the similarly evolved (age $\sim$5 Myr) molecular disk orbiting another X-ray-luminous T Tauri star, T Cha, also appears to show enhanced CN and HCN emission \citep[][note that C$_2$H has yet to be measured for the T Cha disk]{2014A&A...561A..42S}.  The intrinsic X-ray spectral energy distributions of the central stars of the four evolved T Tauri star/disk systems that have been studied in detail thus far (TW Hya, V4046 Sgr, LkCa~15, and T Cha) appear similar to each other, with contributions from both relatively soft ($\sim$3 MK, accretion shock) and somewhat harder ($\sim$10 MK, coronal) components. This combination likely leads to a very large range of X-ray photon penetration depths within the disk \citep{2013ApJ...765....3S}, enhancing molecular ionization and dissociation rates and perhaps increasing the abundances of species, like C$_2$H, that are the dissociation products of more complex organic molecules. These disk irradiation effects then may be reinforced by dust grain surface chemistry processes that favor enhanced production of organic molecules deep within disks \citep[][]{2014A&A...563A..33W}. 

On the other hand, simulations by \citet{2012ApJ...747..114W} indicate that disk CN and C$_2$H abundances are not very sensitive to X-ray irradiation. This raises the possibility that these species are subthermally excited in the disk, as a consequence of their large column densities and (hence) large optical depths (Table~\ref{tab:hfs}); the resulting ``photon trapping'' would effectively lower the critical densities for excitation. Interferometric imaging of the TW Hya and V4046 Sgr disks in lines of CN and C$_2$H that can elucidate the radial and vertical distributions of these species within the disk will be required to distinguish between irradiation and other mechanisms that might enhance their relative abundances.

\section{Conclusions}

We have used the APEX 12 m telescope and broad-band spectrometer to conduct comprehensive mm-wave molecular emission line surveys of the evolved circumstellar disks orbiting the nearby, evolved T Tauri stars TW Hya and V4046~Sgr~AB over the frequency range 275--357 GHz. We find that lines of $^{12}$CO, $^{13}$CO, HCN, CN, and C$_2$H constitute the strongest molecular emission from both disks in this spectral region. The molecule C$_2$H is detected here for the first time in both disks, as is CS in the TW Hya disk. 
The survey results also include the first measurements of the full suite of hyperfine transitions of CN $N=3\rightarrow2$ and C$_2$H $N=4\rightarrow3$ in both disks. Modeling of these CN and C$_2$H hyperfine complexes in the spectrum of TW~Hya indicates that the emission from both species is optically thick and suggests either that emission from CN and C$_2$H originates from very cold ($\stackrel{<}{\sim}$10 K) regions, or that these molecules are subthermally excited. The former possibility, if confirmed, would suggest efficient production of CN and C$_2$H in the outer disk and/or near the disk midplane. It furthermore appears that the fractional abundances of CN and, especially, C$_2$H are significantly enhanced in these evolved protoplanetary disks relative to the fractional abundances of the same molecules in the environments of deeply embedded protostars. 

Clearly, additional work will be necessary to understand the various molecular abundance anomalies apparent in evolved disks orbiting relatively ``old'' T Tauri stars. In this regard, several lines of research are warranted: molecular line surveys of additional disks representing a wider range of evolutionary states; interferometric imaging of the T Tauri star disks in Table~\ref{tbl:ColDensComp} (as well as T Cha) in lines of CN, C$_2$H, and other potential molecular tracers of X-irradiation at high sensitivity and angular resolution (i.e., with ALMA); and detailed numerical simulations of these same systems with an irradiated disk model that incorporates key, observationally-constrained star/disk system parameters --- such as specific incident stellar FUV and X-ray radiation fields, as well as disk masses, radii, gas compositions, gas-to-dust ratios, and viewing angles.

\acknowledgments{\it This publication is based on data acquired with the Atacama Pathfinder Experiment (APEX). APEX is a collaboration between the Max-Planck-Institut fur Radioastronomie, the European Southern Observatory, and the Onsala Space Observatory.
The authors gratefully acknowledge illuminating discussions with Uma Gorti, Karin Oberg, Charlie Qi, and David Wilner, as well as incisive comments from the anonymous referee, all of which significantly improved this paper. This research is supported by National Science Foundation grant AST-1108950 to RIT. D.R.R. acknowledges support from Chilean FONDECYT grant 3130520.}


\newpage

\newpage

\input{Table1rev}

\input{Table2rev}

\input{Table3rev}

\input{Table4}

\input{Table5}

\input{Table6rev}

\end{document}

%% file: Table1rev.tex

\begin{table}[!t]
\caption{\sc TW Hya \& V4046 Sgr: Molecular Species and Transitions
  Detected$^a$ by APEX} 
\begin{center}
\begin{tabular}{cllccccc}
\hline
\hline
Species & Trans. & $\nu$ & $E_{u}/k$ & $A_{ul}$ & $\mu$& \multicolumn{2}{c}{$I$ (Jy km s$^{-1}$)}  \\
                &        & (GHz) & (K) & (s$^{-1}$) & (Debye) & TW Hya & V4046 Sgr \\
\hline
$^{12}$CO & $J=3\rightarrow 2$ & 345.7960 & ...& ...& ... & 21.3 (0.3) &66.3 (2.2) \\
$^{13}$CO & $J=3\rightarrow 2$ & 330.5881 & 31.7319 & 2.199$\times10^{-6}$ & 0.11046& 4.02 (0.40)  & 19.6 (1.4) \\
C$^{18}$O & $J=3\rightarrow 2$ & 329.3305 & ... & ... & ... & 2.0$^b$  & $<$10 \\
C$_2$H & $N=4\rightarrow 3$ & 349.34--349.40  & 41.9148 & 1.411$\times10^{-4}$ & 0.8 & 10.8$^c$   & 22.8$^c$ \\
CN & $N=3\rightarrow 2$ & 340.03--340.26  & 32.6280 & 4.123$\times10^{-4}$  & 1.45  & 51.3$^c$ & 54.0$^c$ \\
HCN & $J=4\rightarrow 3$ & 354.505476 & 42.5348  & 2.047$\times10^{-3}$  & 2.98 & 4.92 (0.81)  & 14.6 (0.8) \\
N$_2$H$^+$ & $J=3\rightarrow 2$ & 279.512 & ...& ...& ... & 2.87 (0.41) &$<$5 \\
DCO$^+$  & $J=4\rightarrow 3$ & 288.144 & ...& ...& ...& 1.64 (0.28) & $<$5 \\
CS &  $J=6\rightarrow 5$ & 293.912244 & 49.3717 & 5.240$\times10^{-4}$
& 1.96 & 2.26 (0.28)  & $<$14 \\ 
       &  $J=7\rightarrow 6$ & 342.88300 & ...& ...& ... & 2.39 (0.31)   &  $<$8 \\
H$^{13}$CO$^+$  & $J=4\rightarrow 3$ & 346.9983 & ...& ...& ... & 1.2$^b$  &   $<$5 \\
\hline
\end{tabular}
\end{center}

a) Upper limits are 3$\sigma$. \\
b) Measured line intensity has a large uncertainty.\\
c) Sum of integrated intensities of hyperfine structure lines
(with estimated uncertainties $\sim$10\%); 
see Table~\ref{tbl:CNandCCHresults}.
\label{tbl:APEXmolecules}
\end{table}%


%% file: Table2rev.tex

%

\begin{deluxetable}{c c c c c c c}

\rotate
\tablecolumns{7}
\tablewidth{0pt}
\tabletypesize{\footnotesize}

\tablecaption{\label{tbl:MolecularSpecies}\sc Molecular Species Detected$^a$ toward TW Hya and/or V4046 Sgr, 215--373 GHz}

\tablehead{
\colhead{} &
\colhead{} &
\colhead{} &
	\multicolumn{2}{c}{TW Hya} &
	\multicolumn{2}{c}{V4046 Sgr} \\
\colhead{Species}&
\colhead{Trans.} &
\colhead{$\nu$} &
\colhead{$I$$^b$} &
\colhead{Refs.$^{c}$} &
\colhead{$I$} &
\colhead{Refs.$^{c}$} \\
\colhead{}&
\colhead{}&
\colhead{(GHz)}&
\colhead{(Jy km s$^{-1}$)}&
\colhead{}&
\colhead{(Jy km s$^{-1}$)}&
\colhead{}
}

\startdata

$^{12}$CO 
          & $J=2\rightarrow 1$ & 230.5380 & 15.9 & 2 & 17.3, 20.0, 34.5 & 7, 8, 9 \\
          & $J=3\rightarrow 2$ & 345.7960 & 21.3, 29.8, 30.9 & 1, 2, 3 & 66.3 & 1 \\
$^{13}$CO & $J=2\rightarrow 1$ & 220.3987 & 2.2, 2.7 & 2, 4 & 9.1, 9.4 & 7, 9 \\
         & $J=3\rightarrow 2$ & 330.5881 & 4.0, 3.7 & 1, 3 & 19.6 & 1\\
C$^{18}$O & $J=2\rightarrow 1$ & 219.560 & 0.68  & 4 & 0.6$^d$ & 9 \\
                & $J=3\rightarrow 2$ & 329.3305 & 2.1$^d$  & 1 & $<$10 & 1 \\
C$_2$H 
       & $N=4\rightarrow 3$ & 349.338  & 5.4$^e$ & 1 & 12.2$^e$ & 1 \\
          & & 349.399 & 5.2$^e$ & 1 & 10.6$^e$ & 1 \\
CN & $N=2\rightarrow 1$ & 226.6794 & \nodata & & 1.29 & 8 \\
       &  & 226.8747 & \nodata & & 17.7, 12.0 & 7, 8 \\
    & $N=3\rightarrow 2$ & 340.0315 & 24.2$^e$, 12.5$^e$  & 1, 2 & 22.6$^e$ & 1 \\
       &  & 340.2478 & 27.1$^e$, 18.7$^e$, 17.8$^e$ & 1, 2, 3 & 31.4$^e$ & 1 \\
HCN & $J=3\rightarrow 2$ & 265.8864 & 7.0 & 2 & 10.3, 9.9 & 7, 8  \\
    & $J=4\rightarrow 3$ & 354.505476 & 4.9, 8.3, 7.6 & 1, 2, 3 & 14.6 & 1 \\
DCN  & $J=3\rightarrow 2$ & 217.2386 & \nodata$^f$ & & $<$0.38 & 8 \\
	   & $J=5\rightarrow 4$ & 362.046 & $<$0.5 & 3 & \nodata &  \\
H$^{13}$CN  & $J=4\rightarrow 3$ & 345.339 & $<$0.9, $<$0.6 & 1, 3 & $<$5 & 1 \\
HNC  & $J=4\rightarrow 3$ & 362.630 & $<$0.8 & 3 & \nodata &  \\
HCO$^{+}$ 
                    & $J=3\rightarrow 2$ & 267.557625 & \nodata$^f$ & & 11.3, 11.43 & 7, 8 \\
	   & $J=4\rightarrow 3$ & 356.73425 & 21.8, 19.7 & 2, 3 & \nodata & \\
DCO$^+$ & $J=3\rightarrow 2$ & 216.1126 & \nodata$^f$ & & 0.80 & 8  \\
        & $J=4\rightarrow 3$ & 288.144 & 1.6 & 1 & $<$5 & 1 \\
        & $J=5\rightarrow 4$ & 360.169 & 1.7 & 3 & \nodata & \\
H$^{13}$CO$^{+}$  & $J=4\rightarrow 3$ & 346.998 & 1.2$^d$, 1.1 & 1, 3 & $<$5 & 1 \\
N$_2$H$^{+}$ & $J=3\rightarrow 2$ & 279.512 & 2.9, 2.6, 2.2 & 1, 4, 5 & $<$5, 2.6 & 1, 8 \\
                      & $J=4\rightarrow 3$ & 372.672 & $<$5 & 3 & \nodata & \\ 
CS &  $J=6\rightarrow 5$ &  293.912244 & 2.3 & 1 & $<$14 & 1 \\
   &  $J=7\rightarrow 6$ &  342.88300 & 2.4 & 1 & $<$8 & 1 \\
H$_{2}$CO 
          & $J=3\rightarrow 2$ & 225.697 & $<$0.8 & 3 & 1.01 & 8 \\
          & $J=4\rightarrow 3$ & 281.5269 & $<$2.5, 1.2 & 1, 5 & $<$5, 0.95 & 1, 8 \\
          & & 300.8366 &$<$2.5 & 1& $<$5 & 1 \\
          & $J=5\rightarrow 4$ & 351.768 & $<$0.6, 0.54 & 3, 5 & $<$5 & 1 \\
\hline
{\it CO$^+$} & $N=3\rightarrow 2$ & 354.0142 & $<$2.5  & 1 & $<$10 & 1 \\
{\it SO}  & $J=8\rightarrow 7$ & 344.310 & $<$2.5, $<$1.6 & 1, 3 & $<$5 & 1 \\
{\it CH$_{3}$OH} & $J=7\rightarrow 6$ & 338.409 & $<$2.5, $<$0.3 & 1, 3 & $<$5 & 1 \\
{\it H$_{2}$D$^{+}$} & $J=1\rightarrow 1$ & 372.421 & $<$1 & 6 & \nodata & \\

\enddata

\vspace{.2in}
a) Selected species with transitions in this range that have not been
detected in either disk are indicated in italics. Ellipses indicate transitions that
are not covered by our line surveys and 
have yet to be measured in one of the two disks. Upper limits are
3$\sigma$ apart from measurements by \citet{2004A&A...425..955T},
which are 2$\sigma$. \\
b) JCMT measurements from \citet{1997Sci...277...67K} and \citet{2004A&A...425..955T}
converted to Jy km s$^{-1}$ assuming a conversion of 15.6 Jy K$^{-1}$
(see http://docs.jach.hawaii.edu/JCMT/HET/GUIDE/het\_guide.ps). \\
c) References: 1. this work; 2. \citet{1997Sci...277...67K};
3. \citet{2004A&A...425..955T}; 4. \citet{2013Sci...341..630Q}; 5. \citet{2013ApJ...765...34Q}
6. \citet{2011A&A...533A.143C}; 7. \citet{2008A&A...492..469K};
8. \citet{2011ApJ...734...98O}; 9. \citet{2013ApJ...775..136R}. \\
d) Measured line intensity is highly uncertain. \\
e)  Integrated line intensities for hyperfine complexes centered at
the listed frequencies 
(see Table~\ref{tbl:CNandCCHresults}). \\
f) Detections of the $J=3\rightarrow 2$ transitions of DCN and DCO$^+$
were reported in \citet{2008ApJ...681.1396Q} and
\citet{2012ApJ...749..162O}; detection of the $J=3\rightarrow 2$
transition of HCN was reported in \citet{2008ApJ...681.1396Q}.
\end{deluxetable}



%% file: Table3rev.tex
\begin{table}[htdp]
\caption{\sc TW Hya \& V4046 Sgr: $^{13}$CO, HCN, and CS Column Densities$^a$} 
\begin{center}
\begin{tabular}{ccccccc}
\hline
\hline
               &            &      & \multicolumn{2}{c}{TW Hya} & \multicolumn{2}{c}{V4046 Sgr} \\
Species (Trans.) & $T_{ex}$ & $\log{Q(T)}$ & $N_X$ & $N_X/N_{^{13}CO}$ & $N_X$ & $N_X/N_{^{13}CO}$ \\
                         &  (K)          &         &  (cm$^{-2}$) & & (cm$^{-2}$) & \\
\hline
$^{13}$CO (3--2) &  37.5   & 1.162 & 2.0$\times10^{14}$ & ... & 9.9$\times10^{14}$ & ... \\
                           &  18.75 & 0.871 & 2.4$\times10^{14}$ & ... & 1.2$\times10^{15}$ & ... \\
                           &   9.375 & 0.589& 6.8$\times10^{14}$ & ... & 3.3$\times10^{15}$ & ... \\
                           &   5.0    & 0.347 & 7.5$\times10^{15}$ & ... & 3.7$\times10^{16}$ & ... \\
HCN (4--3)         &   37.5 & 1.254 & 3.9$\times10^{11}$ & 2.0$\times10^{-3}$ & 1.2$\times10^{12}$ & 1.2$\times10^{-3}$ \\
                           &  18.75 & 0.961 & 6.2$\times10^{11}$ & 2.6$\times10^{-3}$ & 2.0$\times10^{12}$ & 1.7$\times10^{-3}$ \\
                           &   9.375 & 0.676 & 3.1$\times10^{12}$ & 4.6$\times10^{-3}$ & 9.8$\times10^{12}$ & 3.0$\times10^{-3}$ \\
                           &   5.0    & 0.429 & 9.3$\times10^{13}$ & 0.012 & 2.9$\times10^{14}$ & 7.8$\times10^{-3}$ \\
CS (6--5)            &     37.5  & 1.508 & 7.2$\times10^{11}$ & 3.6$\times10^{-3}$ & $<$4.5$\times10^{12}$ & $<$4.5$\times10^{-3}$ \\
                           &  18.75 & 1.212 & 1.4$\times10^{12}$ & 5.8$\times10^{-3}$ & $<$8.4$\times10^{12}$ & $<$7.0$\times10^{-3}$ \\
                           &   9.375 & 0.920 & 9.6$\times10^{12}$ & 0.014 & $<$6.0$\times10^{13}$ & $<$0.18 \\
                           &   5.0    & 0.661 & 5.3$\times10^{14}$ & 0.071 & $<$3.8$\times10^{15}$ &$<$0.10 \\
\hline
\end{tabular}
\end{center}

a) Source-averaged column densities assuming optically thin emission and
adopting source radii (for all molecules) of $5''$ for both
disks. Upper limits for $N_X$ values are based on 3$\sigma$ upper
limits for line intensities. See \S 3.4. \\
\label{tbl:ColumnDensities}
\end{table}%

%% file: Table4.tex

\begin{table}[htdp]
\rotate
\scriptsize
\caption{\sc TW Hya \& V4046 Sgr: Transitions of CN and C$_2$H Measured by APEX} 
\begin{center}
\begin{tabular}{clcccccc}
\hline
\hline
        &   &  &  &  & & TW Hya & V4046 Sgr \\
Species & Transition & $\nu$$^a$ & \aul$^a$ & \gup $^a$ & R.I.$^a$ & \multicolumn{2}{c}{$I$} \\
(Rot. trans.)        &        & (GHz) & (s$^{-1}$) &  &  & \multicolumn{2}{c}{(Jy km s$^{-1}$)} \\
\hline
CN & $J=5/2\rightarrow3/2$, $F=3/2\rightarrow5/2$  & 339.992257 & 3.89$\times10^{-6}$ &  4 & 0.0009 & 0.04 & ... \\
($N=3\rightarrow 2$)  & $J=5/2\rightarrow3/2$, $F=5/2\rightarrow5/2$ & 340.008126 & 
  6.20$\times10^{-5}$ &  6 & 0.0218 & 1.59 & ... \\
  & $J=5/2\rightarrow3/2$, $F=3/2\rightarrow3/2$ & 340.019626 &
  9.27$\times10^{-5}$ &  4 & 0.0217 & 1.10 & ... \\
  & $J=5/2\rightarrow3/2$, $F=7/2\rightarrow5/2$ & 340.031549 &
  3.85$\times10^{-4}$ &  8 & 0.1801 & 10.4 & 22.6$^b$ \\  
  & $J=5/2\rightarrow3/2$, $F=3/2\rightarrow1/2$ & 340.035408 &
  2.89$\times10^{-4}$ &  4 & 0.0676 & 3.11 & ... \\  
  & $J=5/2\rightarrow3/2$, $F=3/2\rightarrow1/2$ & 340.035408 &
  3.23$\times10^{-4}$ &  6 & 0.1135 & 6.60 & ... \\        
  & $J=7/2\rightarrow5/2$, $F=7/2\rightarrow5/2$ & 340.247770 &
  3.80$\times10^{-4}$ &  8 & 0.1778 & 10.3 & 31.6$^c$ \\        
  & $J=7/2\rightarrow5/2$, $F=9/2\rightarrow7/2$ & 340.247770 &
  4.13$\times10^{-4}$ & 10 & 0.2419 & 12.5 & ... \\        
  & $J=7/2\rightarrow5/2$, $F=5/2\rightarrow3/2$ & 340.248544 &
  3.67$\times10^{-4}$ &  6 & 0.1290 & 7.22 & ... \\        
  & $J=7/2\rightarrow5/2$, $F=5/2\rightarrow5/2$ & 340.261773 &
  4.48$\times10^{-5}$ &  6 & 0.0157 & 1.19 & ... \\ 
  & $J=7/2\rightarrow5/2$, $F=7/2\rightarrow7/2$ & 340.264949 &
  3.35$\times10^{-5}$ &  8 & 0.0157 & 1.56 & ... \\ 
  & $J=7/2\rightarrow5/2$, $F=5/2\rightarrow7/2$ & 340.279120  &
 9.27$\times10^{-7}$ &  6 & 0.0003 & 0.04  & ... \\ 
\hline      
C$_2$H & $J=9/2\rightarrow7/2, F=5\rightarrow4$ & 349.337706 & 1.31$\times10^{-4}$ & 11
& 0.3105 & 2.7 & 12$^d$ \\ 
  ($N=4\rightarrow 3$)  & $J=9/2\rightarrow7/2, F=4\rightarrow3$ &
349.338988 & 1.28$\times10^{-4}$ &  9 & 0.2481 & 2.7  & ... \\   
  & $J=7/2\rightarrow5/2, F=4\rightarrow3$ & 349.399276 & 1.25$\times10^{-4}$ &
  9 & 0.2434 & 2.7  & 11$^e$ \\   
  & $J=7/2\rightarrow5/2, F=4\rightarrow3$ & 349.400671 & 1.20$\times10^{-4}$ &
  7 & 0.1816 & 2.5  & ... \\   
\hline
\end{tabular}
\end{center}
a) Values of frequencies, $A_{ul}$, \gup, and theoretical relative
intensities (R.I.) of hyperfine transitions of CN and C$_2$H obtained
from JPL and CDMS databases,
\cite{muller2000}, and \cite{skatrud1983}.\\
b) Sum of integrated intensities of hyperfine structure lines in range
340.032--340.035 GHz.\\
c) Sum of integrated intensities of hyperfine structure lines in range
340.248--340.249 GHz. \\
d) Sum of integrated intensities of hyperfine structure lines in range
349.338--349.339 GHz. \\
e) Sum of integrated intensities of hyperfine structure lines in range
349.399--349.400 GHz.
\label{tbl:CNandCCHresults}
\end{table}%


%% file: Table5.tex
\begin{table}
  \caption{\label{tab:hfs} \sc Results of CN and C$_2$H hyperfine structure analysis for TW Hya}\medskip
  \begin{center}
 \begin{tabular}{c c c c c c c c}
    \hline
   \hline
    Species & $v_0$ & FWHM & $\tau^a$ & $T_{\rm ex}$ & $N_{\rm tot}/\qtot$$^b$ & \qtot($T_{\rm ex}$)$^c$
    & \Ntot $^d$\\
            & (km s$^{-1}$)  & (km s$^{-1}$) &             & (K)     & (cm$^{-2}$)                  &
            & (cm$^{-2}$) \\
    \hline
    CN      & 2.80$\pm$0.10 & 0.61$\pm$0.02 & 4.7$\pm$0.6 & 10.8$\pm$0.8 & 3.7$\times10^{12}$ & 25.9$\pm$1.5
    & (9.6$\pm$1.0)$\times10^{13}$\\
    C$_2$H     & 2.90$\pm$0.05 & 0.48$\pm$0.05 & 12.6$\pm$5.0 & 6.0$\pm$0.9 & 4.0$\times10^{14}$ & 12.9$\pm$1.9
    & (5.1$\pm$3.0)$\times10^{15}$\\
    \hline
  \end{tabular}
\end{center}

\noindent
\footnotesize
a)  Total opacity obtained as the sum of individual hyperfine line opacities $\tau_{ul}$, where $\tau_{ul} = {\rm R.I.}\times \tau$.\\
b) From Eq.~\ref{eq:ntot-hf}. \\
c) From a linear interpolation to tabulated partition function values obtained from the CDMS database.\\
d) Column densities assuming a source radius of $5''$ for both
molecules. See \S 3.4.

\end{table}

%% file: Table6rev.tex

\begin{table}[htdp]
\footnotesize
\caption{\sc Comparison of Fractional Abundances Relative to
  $^{13}$CO, $N$(X)/$N$($^{13}$CO)} 
\begin{center}
\begin{tabular}{cccccccc}
\hline
\hline
{\sc Molecule}               & \multicolumn{4}{c}{\sc  Protoplanetary Disks$^a$} &
               \multicolumn{2}{c}{\sc Protostars$^b$} & PN$^c$\\
& TW Hya & V4046 Sgr & LkCa 15 & DM Tau &   IRS
7B & IRAS 16293 & NGC 7027 \\
\hline
HCN & $2.6\times10^{-3}$ & $1.7\times10^{-3}$ & $5.0\times10^{-3}$ & $1.7\times10^{-3}$
& $6.6\times10^{-4}$ & $1.7\times10^{-3}$ & $1.0\times10^{-4}$ \\
CS & $5.8\times10^{-3}$ & $<7.0\times10^{-3}$ & 0.015 & $1.0\times10^{-3}$ & $1.7\times10^{-3}$
& $4.6\times10^{-3}$ & ... \\
CN & 0.40 & 0.08$^d$ & 0.036 & 0.01 & $1.0\times10^{-3}$ &
$1.2\times10^{-4}$ & $3.0\times10^{-3}$ \\
C$_2$H & 21 & 8$^d$ & 0.16 & 0.03 & $3.8\times10^{-3}$ &
$2.0\times10^{-4}$ & $2.0\times10^{-3}$\\
\hline
\end{tabular}
\end{center}

a) Based on results listed in Tables \ref{tbl:ColumnDensities} and
\ref{tab:hfs}; \citet[][their Table 3]{2010ApJ...714.1511H};
\citet[][their Table 9]{2004A&A...425..955T}; and \citet[][their Table 1 and
refs.\ therein]{1997A&A...317L..55D}. See \S 4.\\
b) R CrA IRS 7B relative abundances from \citet[][their Table 4,
and refs.\ therein]{2012ApJ...745..126W}, assuming $N$(H$_2$)/$N$($^{13}$CO)
$=6\times10^5$; 
IRAS 16293--2422 relative abundances from
\citet[][their Table 5]{2002A&A...390.1001S}. \\
c) Relative abundances for the planetary nebula (PN) NGC 7027 from
\citet[][their Table 3]{2001ApJ...562..824H}. \\
d) Estimated abundance ratio is highly uncertain.

\label{tbl:ColDensComp}
\end{table}%

%% file: LineSurveysFullPostProofsText.bbl
\begin{thebibliography}{71}
\expandafter\ifx\csname natexlab\endcsname\relax\def\natexlab#1{#1}\fi

\bibitem[{{Andrews} {et~al.}(2012){Andrews}, {Wilner}, {Hughes}, {Qi},
  {Rosenfeld}, {{\"O}berg}, {Birnstiel}, {Espaillat}, {Cieza}, {Williams},
  {Lin}, \& {Ho}}]{2012ApJ...744..162A}
{Andrews}, S.~M., {et~al.} 2012, \apj, 744, 162

\bibitem[{{Aresu} {et~al.}(2012){Aresu}, {Meijerink}, {Kamp}, {Spaans}, {Thi},
  \& {Woitke}}]{2012A&A...547A..69A}
{Aresu}, G., {Meijerink}, R., {Kamp}, I., {Spaans}, M., {Thi}, W.-F., \&
  {Woitke}, P. 2012, \aap, 547, A69

\bibitem[{{Bergin} {et~al.}(2013){Bergin}, {Cleeves}, {Gorti}, {Zhang},
  {Blake}, {Green}, {Andrews}, {Evans}, {Henning}, {{\"O}berg}, {Pontoppidan},
  {Qi}, {Salyk}, \& {van Dishoeck}}]{2013Natur.493..644B}
{Bergin}, E.~A., {et~al.} 2013, \nat, 493, 644

\bibitem[{{Binks} \& {Jeffries}(2014)}]{2014MNRAS.438L..11B}
{Binks}, A.~S., \& {Jeffries}, R.~D. 2014, \mnras, 438, L11

\bibitem[{{Blake} {et~al.}(1986){Blake}, {Masson}, {Phillips}, \&
  {Sutton}}]{1986ApJS...60..357B}
{Blake}, G.~A., {Masson}, C.~R., {Phillips}, T.~G., \& {Sutton}, E.~C. 1986,
  \apjs, 60, 357

\bibitem[{{Blake} {et~al.}(1994){Blake}, {van Dishoeck}, {Jansen}, {Groesbeck},
  \& {Mundy}}]{1994ApJ...428..680B}
{Blake}, G.~A., {van Dishoeck}, E.~F., {Jansen}, D.~J., {Groesbeck}, T.~D., \&
  {Mundy}, L.~G. 1994, \apj, 428, 680

\bibitem[{{Bottinelli} {et~al.}(2007){Bottinelli}, {Ceccarelli}, {Williams}, \&
  {Lefloch}}]{2007A&A...463..601B}
{Bottinelli}, S., {Ceccarelli}, C., {Williams}, J.~P., \& {Lefloch}, B. 2007,
  \aap, 463, 601

\bibitem[{{Caux} {et~al.}(2011){Caux}, {Kahane}, {Castets}, {Coutens},
  {Ceccarelli}, {Bacmann}, {Bisschop}, {Bottinelli}, {Comito}, {Helmich},
  {Lefloch}, {Parise}, {Schilke}, {Tielens}, {van Dishoeck}, {Vastel},
  {Wakelam}, \& {Walters}}]{2011A&A...532A..23C}
{Caux}, E., {et~al.} 2011, \aap, 532, A23

\bibitem[{{Chapillon} {et~al.}(2012){Chapillon}, {Guilloteau}, {Dutrey},
  {Pi{\'e}tu}, \& {Gu{\'e}lin}}]{2012A&A...537A..60C}
{Chapillon}, E., {Guilloteau}, S., {Dutrey}, A., {Pi{\'e}tu}, V., \&
  {Gu{\'e}lin}, M. 2012, \aap, 537, A60

\bibitem[{{Chapillon} {et~al.}(2011){Chapillon}, {Parise}, {Guilloteau}, \&
  {Du}}]{2011A&A...533A.143C}
{Chapillon}, E., {Parise}, B., {Guilloteau}, S., \& {Du}, F. 2011, \aap, 533,
  A143

\bibitem[{{Cleeves} {et~al.}(2013){Cleeves}, {Adams}, \&
  {Bergin}}]{2013ApJ...772....5C}
{Cleeves}, L.~I., {Adams}, F.~C., \& {Bergin}, E.~A. 2013, \apj, 772, 5

\bibitem[{{Drake} {et~al.}(2009){Drake}, {Ercolano}, {Flaccomio}, \&
  {Micela}}]{2009ApJ...699L..35D}
{Drake}, J.~J., {Ercolano}, B., {Flaccomio}, E., \& {Micela}, G. 2009, \apjl,
  699, L35

\bibitem[{{Ducourant} {et~al.}(2014){Ducourant}, {Teixeira}, {Galli}, {Le
  Campion}, {Krone-Martins}, {Zuckerman}, {Chauvin}, \&
  {Song}}]{2014A&A...563A.121D}
{Ducourant}, C., {Teixeira}, R., {Galli}, P.~A.~B., {Le Campion}, J.~F.,
  {Krone-Martins}, A., {Zuckerman}, B., {Chauvin}, G., \& {Song}, I. 2014,
  \aap, 563, A121

\bibitem[{{Dutrey} {et~al.}(1997){Dutrey}, {Guilloteau}, \&
  {Guelin}}]{1997A&A...317L..55D}
{Dutrey}, A., {Guilloteau}, S., \& {Guelin}, M. 1997, \aap, 317, L55

\bibitem[{{Dutrey} {et~al.}(2014){Dutrey}, {Semenov}, {Chapillon}, {Gorti},
  {Guilloteau}, {Hersant}, {Hogerheijde}, {Hughes}, {Meeus}, {Nomura},
  {Pi{\'e}tu}, {Qi}, \& {Wakelam}}]{2014arXiv1402.3503D}
{Dutrey}, A., {et~al.} 2014, ArXiv e-prints

\bibitem[{{Ehrenfreund} \& {Charnley}(2000)}]{2000ARA&A..38..427E}
{Ehrenfreund}, P., \& {Charnley}, S.~B. 2000, \araa, 38, 427

\bibitem[{{Ercolano} {et~al.}(2009){Ercolano}, {Clarke}, \&
  {Drake}}]{2009ApJ...699.1639E}
{Ercolano}, B., {Clarke}, C.~J., \& {Drake}, J.~J. 2009, \apj, 699, 1639

\bibitem[{{Ercolano} {et~al.}(2008){Ercolano}, {Drake}, {Raymond}, \&
  {Clarke}}]{2008ApJ...688..398E}
{Ercolano}, B., {Drake}, J.~J., {Raymond}, J.~C., \& {Clarke}, C.~C. 2008,
  \apj, 688, 398

\bibitem[{{Glassgold} {et~al.}(2012){Glassgold}, {Galli}, \&
  {Padovani}}]{2012ApJ...756..157G}
{Glassgold}, A.~E., {Galli}, D., \& {Padovani}, M. 2012, \apj, 756, 157

\bibitem[{{Glassgold} {et~al.}(2009){Glassgold}, {Meijerink}, \&
  {Najita}}]{2009ApJ...701..142G}
{Glassgold}, A.~E., {Meijerink}, R., \& {Najita}, J.~R. 2009, \apj, 701, 142

\bibitem[{{Glassgold} {et~al.}(1997){Glassgold}, {Najita}, \&
  {Igea}}]{1997ApJ...480..344G}
{Glassgold}, A.~E., {Najita}, J., \& {Igea}, J. 1997, \apj, 480, 344

\bibitem[{{Glassgold} {et~al.}(2004){Glassgold}, {Najita}, \&
  {Igea}}]{2004ApJ...615..972G}
---. 2004, \apj, 615, 972

\bibitem[{{Goldsmith} \& {Langer}(1999)}]{1999ApJ...517..209G}
{Goldsmith}, P.~F., \& {Langer}, W.~D. 1999, \apj, 517, 209

\bibitem[{{Gorti} {et~al.}(2009){Gorti}, {Dullemond}, \&
  {Hollenbach}}]{2009ApJ...705.1237G}
{Gorti}, U., {Dullemond}, C.~P., \& {Hollenbach}, D. 2009, \apj, 705, 1237

\bibitem[{{Gorti} \& {Hollenbach}(2008)}]{2008ApJ...683..287G}
{Gorti}, U., \& {Hollenbach}, D. 2008, \apj, 683, 287

\bibitem[{{Gorti} {et~al.}(2011){Gorti}, {Hollenbach}, {Najita}, \&
  {Pascucci}}]{2011ApJ...735...90G}
{Gorti}, U., {Hollenbach}, D., {Najita}, J., \& {Pascucci}, I. 2011, \apj, 735,
  90

\bibitem[{{G{\"u}nther} {et~al.}(2006){G{\"u}nther}, {Liefke}, {Schmitt},
  {Robrade}, \& {Ness}}]{2006A&A...459L..29G}
{G{\"u}nther}, H.~M., {Liefke}, C., {Schmitt}, J.~H.~M.~M., {Robrade}, J., \&
  {Ness}, J.-U. 2006, \aap, 459, L29

\bibitem[{{Hasegawa} \& {Kwok}(2001)}]{2001ApJ...562..824H}
{Hasegawa}, T.~I., \& {Kwok}, S. 2001, \apj, 562, 824

\bibitem[{{Henning} {et~al.}(2010){Henning}, {Semenov}, {Guilloteau}, {Dutrey},
  {Hersant}, {Wakelam}, {Chapillon}, {Launhardt}, {Pi{\'e}tu}, \&
  {Schreyer}}]{2010ApJ...714.1511H}
{Henning}, T., {et~al.} 2010, \apj, 714, 1511

\bibitem[{{Hubbard} {et~al.}(2002){Hubbard}, {Burrows}, \&
  {Lunine}}]{2002ARA&A..40..103H}
{Hubbard}, W.~B., {Burrows}, A., \& {Lunine}, J.~I. 2002, \araa, 40, 103

\bibitem[{{Johansson} {et~al.}(1984){Johansson}, {Andersson}, {Ellder},
  {Friberg}, {Hjalmarson}, {Hoglund}, {Irvine}, {Olofsson}, \&
  {Rydbeck}}]{1984A&A...130..227J}
{Johansson}, L.~E.~B., {et~al.} 1984, \aap, 130, 227

\bibitem[{{Kaifu} {et~al.}(2004){Kaifu}, {Ohishi}, {Kawaguchi}, {Saito},
  {Yamamoto}, {Miyaji}, {Miyazawa}, {Ishikawa}, {Noumaru}, {Harasawa}, {Okuda},
  \& {Suzuki}}]{2004PASJ...56...69K}
{Kaifu}, N., {et~al.} 2004, \pasj, 56, 69

\bibitem[{{Kastner} {et~al.}(2002){Kastner}, {Huenemoerder}, {Schulz},
  {Canizares}, \& {Weintraub}}]{2002ApJ...567..434K}
{Kastner}, J.~H., {Huenemoerder}, D.~P., {Schulz}, N.~S., {Canizares}, C.~R.,
  \& {Weintraub}, D.~A. 2002, \apj, 567, 434

\bibitem[{{Kastner} {et~al.}(2008){Kastner}, {Zuckerman}, {Hily-Blant}, \&
  {Forveille}}]{2008A&A...492..469K}
{Kastner}, J.~H., {Zuckerman}, B., {Hily-Blant}, P., \& {Forveille}, T. 2008,
  \aap, 492, 469

\bibitem[{{Kastner} {et~al.}(1997){Kastner}, {Zuckerman}, {Weintraub}, \&
  {Forveille}}]{1997Sci...277...67K}
{Kastner}, J.~H., {Zuckerman}, B., {Weintraub}, D.~A., \& {Forveille}, T. 1997,
  Science, 277, 67

\bibitem[{{Kastner} {et~al.}(2012){Kastner}, {Montez}, {Balick}, {Frew},
  {Miszalski}, {Sahai}, {Blackman}, {Chu}, {De Marco}, {Frank}, {Guerrero},
  {Lopez}, {Rapson}, {Zijlstra}, {Behar}, {Bujarrabal}, {Corradi}, {Nordhaus},
  {Parker}, {Sandin}, {Sch{\"o}nberner}, {Soker}, {Sokoloski}, {Steffen},
  {Ueta}, \& {Villaver}}]{2012AJ....144...58K}
{Kastner}, J.~H., {et~al.} 2012, \aj, 144, 58

\bibitem[{{Klein} {et~al.}(2006){Klein}, {Philipp}, {Kr{\"a}mer}, {Kasemann},
  {G{\"u}sten}, \& {Menten}}]{klein2006}
{Klein}, B., {Philipp}, S.~D., {Kr{\"a}mer}, I., {Kasemann}, C., {G{\"u}sten},
  R., \& {Menten}, K.~M. 2006, \aap, 454, L29

\bibitem[{{Luu} \& {Jewitt}(2002)}]{2002ARA&A..40...63L}
{Luu}, J.~X., \& {Jewitt}, D.~C. 2002, \araa, 40, 63

\bibitem[{{Meijerink} {et~al.}(2008){Meijerink}, {Glassgold}, \&
  {Najita}}]{2008ApJ...676..518M}
{Meijerink}, R., {Glassgold}, A.~E., \& {Najita}, J.~R. 2008, \apj, 676, 518

\bibitem[{{Milam} {et~al.}(2005){Milam}, {Savage}, {Brewster}, {Ziurys}, \&
  {Wyckoff}}]{2005ApJ...634.1126M}
{Milam}, S.~N., {Savage}, C., {Brewster}, M.~A., {Ziurys}, L.~M., \& {Wyckoff},
  S. 2005, \apj, 634, 1126

\bibitem[{{M{\"u}ller} {et~al.}(2000){M{\"u}ller}, {Klaus}, \&
  {Winnewisser}}]{muller2000}
{M{\"u}ller}, H.~S.~P., {Klaus}, T., \& {Winnewisser}, G. 2000, \aap, 357, L65

\bibitem[{{{\"O}berg} {et~al.}(2012){{\"O}berg}, {Qi}, {Wilner}, \&
  {Hogerheijde}}]{2012ApJ...749..162O}
{{\"O}berg}, K.~I., {Qi}, C., {Wilner}, D.~J., \& {Hogerheijde}, M.~R. 2012,
  \apj, 749, 162

\bibitem[{{{\"O}berg} {et~al.}(2011){{\"O}berg}, {Qi}, {Fogel}, {Bergin},
  {Andrews}, {Espaillat}, {Wilner}, {Pascucci}, \&
  {Kastner}}]{2011ApJ...734...98O}
{{\"O}berg}, K.~I., {et~al.} 2011, \apj, 734, 98

\bibitem[{{Owen} {et~al.}(2011){Owen}, {Ercolano}, \&
  {Clarke}}]{2011MNRAS.412...13O}
{Owen}, J.~E., {Ercolano}, B., \& {Clarke}, C.~J. 2011, \mnras, 412, 13

\bibitem[{{Owen} {et~al.}(2010){Owen}, {Ercolano}, {Clarke}, \&
  {Alexander}}]{2010MNRAS.401.1415O}
{Owen}, J.~E., {Ercolano}, B., {Clarke}, C.~J., \& {Alexander}, R.~D. 2010,
  \mnras, 401, 1415

\bibitem[{{Pety}(2005)}]{pety2005}
{Pety}, J. 2005, in SF2A-2005: Semaine de l'Astrophysique Francaise, ed.
  F.~{Casoli}, T.~{Contini}, J.~M. {Hameury}, \& L.~{Pagani}, 721

\bibitem[{{Qi} {et~al.}(2013{\natexlab{a}}){Qi}, {{\"O}berg}, \&
  {Wilner}}]{2013ApJ...765...34Q}
{Qi}, C., {{\"O}berg}, K.~I., \& {Wilner}, D.~J. 2013{\natexlab{a}}, \apj, 765,
  34

\bibitem[{{Qi} {et~al.}(2008){Qi}, {Wilner}, {Aikawa}, {Blake}, \&
  {Hogerheijde}}]{2008ApJ...681.1396Q}
{Qi}, C., {Wilner}, D.~J., {Aikawa}, Y., {Blake}, G.~A., \& {Hogerheijde},
  M.~R. 2008, \apj, 681, 1396

\bibitem[{{Qi} {et~al.}(2006){Qi}, {Wilner}, {Calvet}, {Bourke}, {Blake},
  {Hogerheijde}, {Ho}, \& {Bergin}}]{2006ApJ...636L.157Q}
{Qi}, C., {Wilner}, D.~J., {Calvet}, N., {Bourke}, T.~L., {Blake}, G.~A.,
  {Hogerheijde}, M.~R., {Ho}, P.~T.~P., \& {Bergin}, E. 2006, \apjl, 636, L157

\bibitem[{{Qi} {et~al.}(2004){Qi}, {Ho}, {Wilner}, {Takakuwa}, {Hirano},
  {Ohashi}, {Bourke}, {Zhang}, {Blake}, {Hogerheijde}, {Saito}, {Choi}, \&
  {Yang}}]{2004ApJ...616L..11Q}
{Qi}, C., {et~al.} 2004, \apjl, 616, L11

\bibitem[{{Qi} {et~al.}(2013{\natexlab{b}}){Qi}, {{\"O}berg}, {Wilner},
  {D'Alessio}, {Bergin}, {Andrews}, {Blake}, {Hogerheijde}, \& {van
  Dishoeck}}]{2013Sci...341..630Q}
---. 2013{\natexlab{b}}, Science, 341, 630

\bibitem[{{Rodriguez} {et~al.}(2010){Rodriguez}, {Kastner}, {Wilner}, \&
  {Qi}}]{2010ApJ...720.1684R}
{Rodriguez}, D.~R., {Kastner}, J.~H., {Wilner}, D., \& {Qi}, C. 2010, \apj,
  720, 1684

\bibitem[{{Rosenfeld} {et~al.}(2013){Rosenfeld}, {Andrews}, {Wilner},
  {Kastner}, \& {McClure}}]{2013ApJ...775..136R}
{Rosenfeld}, K.~A., {Andrews}, S.~M., {Wilner}, D.~J., {Kastner}, J.~H., \&
  {McClure}, M.~K. 2013, \apj, 775, 136

\bibitem[{{Rosenfeld} {et~al.}(2012{\natexlab{a}}){Rosenfeld}, {Andrews},
  {Wilner}, \& {Stempels}}]{2012ApJ...759..119R}
{Rosenfeld}, K.~A., {Andrews}, S.~M., {Wilner}, D.~J., \& {Stempels}, H.~C.
  2012{\natexlab{a}}, \apj, 759, 119

\bibitem[{{Rosenfeld} {et~al.}(2012{\natexlab{b}}){Rosenfeld}, {Qi}, {Andrews},
  {Wilner}, {Corder}, {Dullemond}, {Lin}, {Hughes}, {D'Alessio}, \&
  {Ho}}]{2012ApJ...757..129R}
{Rosenfeld}, K.~A., {et~al.} 2012{\natexlab{b}}, \apj, 757, 129

\bibitem[{{Sacco} {et~al.}(2014){Sacco}, {Kastner}, {Forveille}, {Principe},
  {Montez}, {Zuckerman}, \& {Hily-Blant}}]{2014A&A...561A..42S}
{Sacco}, G.~G., {Kastner}, J.~H., {Forveille}, T., {Principe}, D., {Montez},
  R., {Zuckerman}, B., \& {Hily-Blant}, P. 2014, \aap, 561, A42

\bibitem[{{Salter} {et~al.}(2011){Salter}, {Hogerheijde}, {van der Burg},
  {Kristensen}, \& {Brinch}}]{2011A&A...536A..80S}
{Salter}, D.~M., {Hogerheijde}, M.~R., {van der Burg}, R.~F.~J., {Kristensen},
  L.~E., \& {Brinch}, C. 2011, \aap, 536, A80

\bibitem[{{Schilke} {et~al.}(1997){Schilke}, {Groesbeck}, {Blake}, \&
  {Phillips}}]{1997ApJS..108..301S}
{Schilke}, P., {Groesbeck}, T.~D., {Blake}, G.~A., \& {Phillips}, T.~G. 1997,
  \apjs, 108, 301

\bibitem[{{Sch{\"o}ier} {et~al.}(2002){Sch{\"o}ier}, {J{\o}rgensen}, {van
  Dishoeck}, \& {Blake}}]{2002A&A...390.1001S}
{Sch{\"o}ier}, F.~L., {J{\o}rgensen}, J.~K., {van Dishoeck}, E.~F., \& {Blake},
  G.~A. 2002, \aap, 390, 1001

\bibitem[{{Scott} {et~al.}(2006){Scott}, {Asplund}, {Grevesse}, \&
  {Sauval}}]{2006A&A...456..675S}
{Scott}, P.~C., {Asplund}, M., {Grevesse}, N., \& {Sauval}, A.~J. 2006, \aap,
  456, 675

\bibitem[{{Skatrud} {et~al.}(1983){Skatrud}, {de Lucia}, {Blake}, \&
  {Sastry}}]{skatrud1983}
{Skatrud}, D.~D., {de Lucia}, F.~C., {Blake}, G.~A., \& {Sastry}, K.~V.~L.~N.
  1983, Journal of Molecular Spectroscopy, 99, 35

\bibitem[{{Skinner} \& {G{\"u}del}(2013)}]{2013ApJ...765....3S}
{Skinner}, S.~L., \& {G{\"u}del}, M. 2013, \apj, 765, 3

\bibitem[{{St{\"a}uber} {et~al.}(2005){St{\"a}uber}, {Doty}, {van Dishoeck}, \&
  {Benz}}]{2005A&A...440..949S}
{St{\"a}uber}, P., {Doty}, S.~D., {van Dishoeck}, E.~F., \& {Benz}, A.~O. 2005,
  \aap, 440, 949

\bibitem[{{Thi} {et~al.}(2004){Thi}, {van Zadelhoff}, \& {van
  Dishoeck}}]{2004A&A...425..955T}
{Thi}, W., {van Zadelhoff}, G., \& {van Dishoeck}, E.~F. 2004, \aap, 425, 955

\bibitem[{{Torres} {et~al.}(2008){Torres}, {Quast}, {Melo}, \&
  {Sterzik}}]{2008hsf2.book..757T}
{Torres}, C.~A.~O., {Quast}, G.~R., {Melo}, C.~H.~F., \& {Sterzik}, M.~F. 2008,
  {Young Nearby Loose Associations}, ed. {Reipurth, B.}, 757--+

\bibitem[{{van Dishoeck} {et~al.}(1995){van Dishoeck}, {Blake}, {Jansen}, \&
  {Groesbeck}}]{1995ApJ...447..760V}
{van Dishoeck}, E.~F., {Blake}, G.~A., {Jansen}, D.~J., \& {Groesbeck}, T.~D.
  1995, \apj, 447, 760

\bibitem[{{Walsh} {et~al.}(2014){Walsh}, {Millar}, {Nomura}, {Herbst}, {Widicus
  Weaver}, {Aikawa}, {Laas}, \& {Vasyunin}}]{2014A&A...563A..33W}
{Walsh}, C., {Millar}, T.~J., {Nomura}, H., {Herbst}, E., {Widicus Weaver}, S.,
  {Aikawa}, Y., {Laas}, J.~C., \& {Vasyunin}, A.~I. 2014, \aap, 563, A33

\bibitem[{{Walsh} {et~al.}(2012){Walsh}, {Nomura}, {Millar}, \&
  {Aikawa}}]{2012ApJ...747..114W}
{Walsh}, C., {Nomura}, H., {Millar}, T.~J., \& {Aikawa}, Y. 2012, \apj, 747,
  114

\bibitem[{{Watanabe} {et~al.}(2012){Watanabe}, {Sakai}, {Lindberg},
  {J{\o}rgensen}, {Bisschop}, \& {Yamamoto}}]{2012ApJ...745..126W}
{Watanabe}, Y., {Sakai}, N., {Lindberg}, J.~E., {J{\o}rgensen}, J.~K.,
  {Bisschop}, S.~E., \& {Yamamoto}, S. 2012, \apj, 745, 126

\bibitem[{{Williams} \& {Cieza}(2011)}]{2011ARA&A..49...67W}
{Williams}, J.~P., \& {Cieza}, L.~A. 2011, \araa, 49, 67

\bibitem[{{Ziurys} {et~al.}(1982){Ziurys}, {Saykally}, {Plambeck}, \&
  {Erickson}}]{1982ApJ...254...94Z}
{Ziurys}, L.~M., {Saykally}, R.~J., {Plambeck}, R.~L., \& {Erickson}, N.~R.
  1982, \apj, 254, 94

\end{thebibliography}
